\documentclass[acmsmall]{acmart}
\renewcommand\footnotetextcopyrightpermission[1]{} 
\AtBeginDocument{%
  \providecommand\BibTeX{{%
    \normalfont B\kern-0.5em{\scshape i\kern-0.25em b}\kern-0.8em\TeX}}}

\usepackage{graphicx}
\usepackage{algorithm}
\usepackage[noend]{algpseudocode}
\usepackage{amsmath}
\usepackage{paralist}

\usepackage{breqn}
\usepackage{subcaption}
\usepackage{multirow}
\usepackage{psfrag}
\usepackage{url}
\usepackage{hyperref}
\usepackage{cleveref}
\usepackage{booktabs}
\usepackage{rotating}
\usepackage{colortbl}
\usepackage{paralist}
\usepackage{epstopdf}
\usepackage{nag}
\usepackage{microtype}
\usepackage{siunitx}
\usepackage{nicefrac}
\usepackage{lipsum}
\usepackage[english]{babel}
\usepackage[pangram]{blindtext}
\usepackage{tikz}
\usepackage{tabularx}
\usetikzlibrary{fit,positioning,arrows.meta,shapes,arrows}
\begin{document}


\title[A Survey of Distributed Ledger Technology for IoT Verticals]{A Survey of Distributed Ledger Technology for IoT Verticals}

\author{Rongxin Xu}
\affiliation{%
  \institution{Hunan Key Laboratory of Data Science \& Blockchain, Business School,
Hunan University}
  \city{Changsha 410082}
  \country{China}
  }

\author{Qiujun Lan}
\affiliation{%
  \institution{Hunan Key Laboratory of Data Science \& Blockchain, Business School,
Hunan University}
  \city{Changsha 410082}
  \country{China}
  }
\email{lanqiujun@hnu.edu.cn}
\orcid{0000-0001-7523-9487}

\author{Shiva Raj Pokhrel}
\authornote{Corresponding Author}
\affiliation{%
  \institution{School of IT,  Deakin University}
  \city{Geelong, VIC 3216}
  \country{Australia}}
\email{shiva.pokhrel@deakin.edu.au}
\orcid{0000-0001-5819-765X}

\author{Gang Li}
\affiliation{%
  \institution{Centre for Cyber Security Research and Innovation, 
Deakin University}
  \city{Geelong, VIC 3216}
  \country{Australia}}
\email{gang.li@deakin.edu.au}
\orcid{0000-0003-1583-641X}


\renewcommand{\shortauthors}{Q. Lan et al.}


\begin{abstract}
	The Internet of Things (IoT) and Distributed ledger technology (DLT)  have significantly changed our daily lives. Due to their distributed operational environment and naturally decentralized applications, the convergence of these two technologies indicates a more lavish arrangement for the future. This article develops a comprehensive survey to investigate and illustrate state-of-the-art DLT for various IoT use cases, from smart homes to autonomous vehicles and smart cities. We develop a novel framework for conducting a systematic and comprehensive review of DLT over IoT by extending the knowledge graph approach. With relevant insights from this review, we extract innovative and pragmatic techniques to DLT design that enable high-performance, sustainable, and highly scalable IoT systems. Our findings support designing an end-to-end IoT-native DLT architecture for the future that fully coordinates network-assisted functionalities.
\end{abstract}



\maketitle


\section{Introduction}\label{sec-intro}

\emph{Distributed Ledger Technology} (DLT)~\cite{rauchsDistributedLedgerTechnology2018}
has started transforming all aspects of modern trade. 
As reported by Tapscott~\citep{tapscottHowBlockchainWill2017}, 
blockchain appears to be a more prominent arrangement 
than the Internet itself
(see~\cite{unalanDemocratisingSystemsInnovations2020} and the references therein).
Consolidating DLT and \emph{Internet of Things} (IoT)~\cite{ashtonThatInternetThings2009}
means that we have two great arrangements together,
and the management of the convergent DLT-IoT framework 
is more challenging and non-trivial.
However, such a convergence is important and will be highly impactful 
as both DLT and IoT are distributed by nature and may complement each other. 

With the rapid advances of the fifth-generation (5G) 
and beyond wireless communication technology, 
IoT has evolved from an early concept~\cite{ashtonThatInternetThings2009}
to a reality~\cite{zhengIoTSecurityAutomation2020}. 
It is well-known that DLTs bring a range of benefits,
which are not only the solutions to existing IoT problems, 
but highly likely to benefit emerging issues of future IoT applications. 
As the IoT systems are distributed  by nature~\cite{jainCloudEdgeDistributed2017}, 
it is expected that DLT, 
such as blockchain,
could play a vital role in orchestrating how the things, 
devices and machines collaborate in a decentralized fashion~\cite{pena-lopezITUInternetReport2005}. 

In this context, 
DLTs are being developed as a basis for applications and devices 
that involve transactions and interactions~\cite{liuDistributedLedgerTechnology2020}.
DLTs have the potential to improve the transactions and interactions among IoT devices.
Moreover, 
DLTs can flexibly integrate IoT contracts
with other intelligent applications and devices (without a third party),
thus explicitly bolstering the internal mechanism of IoT system. 
Data governance in IoT~\cite{sicariDataGovernanceInternet2018}
has usually been a theoretical-sounding concept until recently, 
but it is gradually commencing a real need~\cite{deprieelleRoleEcosystemData2020}.
In addition to the new data privacy regulations, 
such as the \emph{General Data Protection Regulation}~\cite{RegulationEU20162016} 
and the \emph{California Consumer Privacy Act}~\cite{CaliforniaConsumerPrivacy2022}, 
the occurrence of numerous data breaches~\footnote{\url{https://www.csoonline.com/article/2130877/the-biggest-data-breaches-of-the-21st-century.html}} 
and ransomware~\cite{kumarAnatomyRansomwareAttack2021} in IoT 
requires a new data governance approach 
to restoring public trust~\cite{liuSecureEfficientDistributed2020}. 
We advocate that DLTs could help reinstate the trust in IoT applications.

As mentioned earlier,
the gradual convergence of IoT and DLT will
be a greater framework~\cite{farahaniConvergenceIoTDistributed2021}.
In building up and managing the IoT systems, 
DLTs have the potential for its orchestration and automation, 
such as transparency in supply chain initiatives to
encourage self-execution of payments and
enhancing data protection for the distributed processing among 
devices and things~\cite{ahiRoleDistributedLedger2019}.
We observe that 
DLT vows to be a missing link of IoT that 
could empower peer-to-peer negotiated conduct 
between devices with no other party to ``certify'' IoT transactions. 
DLT can address the impending challenges of scalability, 
the primary objective of disappointment between IoT devices and parties, 
and their security, trust, time stamping, record, 
and reliability for efficiency and consistency~\cite{farahaniConvergenceIoTDistributed2021}. 
This is reflected, for example, 
in the adoption of group signatures and blockchain 
by~\citet{xuBlockchainEnabledAccountabilityMechanism2021}  
for anonymous identity management in the smart home IoT applications. 
~\citet{ouaddahNovelPrivacyPreservingAccess2017,ouaddahFairAccessNewBlockchainbased2016} 
developed a DLT access control framework -- \textbf{FairAccess}, 
which treats the access rights as assets 
in the blockchain to replace cryptocurrency 
and exchange the access rights securely through transactions. 
Such a framework can be extended and operationalized for IoT applications.

Some existing platforms and giant companies 
have already considered the potential of this convergence.
IBM Blockchain~\footnote{\url{https://www.ibm.com/thought-leadership/institute-business-value/report/device-democracy}}, 
for example, 
now allows the extension of blockchain to the cognitive IoT. 
Other platforms such as \emph{Amazon Web services} (AWS) IoT, 
\emph{Google Cloud IoT platform} (GCP), 
\emph{Microsoft Azure IoT Suite}, 
and \emph{Oracle IoT} already have DLTs 
in their roadmap~\cite{raySurveyIoTCloud2016}. 
\textbf{IOTA}~\cite{popovTangle2018} 
is another important initiative~\footnote{\url{https://www.iota.org/}}, 
which develops an open-source DLT and cryptocurrency for IoTs. 
We believe that 
the blend of IoT and DLT will generate real momentum across organizations 
and in multiple facets of IoT implementations~\cite{farahaniConvergenceIoTDistributed2021}. 

With DLT-IoT convergence, 
we anticipate major contribution to the evolving digital ecosystem 
that is motivating countless technologies and impacting endless applications, 
from analytics, security, 
privacy to their automation and orchestration~\cite{zhengIoTSecurityAutomation2020}. 
In a distributed IoT environment, 
all aforementioned facets have so far always 
been considered in a centralized fashion,
which potentially causes the \emph{Single Point of Failure} (SPoF)~\cite{romanFeaturesChallengesSecurity2013}).

\subsection{Motivation and Relevant Works}

Fortunately, 
just as IoT devices cross the \num{30} billion imprint~\cite{nordrumPopularInternetThings2016}, 
cutting-edge DLT initiatives such as \textbf{IOTA}~\cite{popovTangle2018}, 
\textbf{Hashgraph}~\cite{bairdSwirldsHashgraphConsensus2016}, 
\textbf{Neural Ledger}~\cite{velascoNeuralDistributedLedger2020} 
and \textbf{Interledger}~\cite{sirisInterledgerApproaches2019} are now on the market. 
These new DLT initiatives exemplify 
attractive designs for the transaction dynamics  
required to support various IoT scenarios (discussed later in \Cref{sec-usecases}). 
Such a blend of IoT with DLT could 
fundamentally transform inter-organizational transactions 
and enable seamless enterprise collaborations. 
However, 
to realize the aforementioned changes and 
the seamless integration of DLT with IoT, 
there are a lot of challenges to be resolved.
We have identified the following key bottleneck challenges, 
which will be discussed in details later in~\Cref{sec-challenges}.

\begin{description}
	\item[Resource limitations.]
	Blockchain, for example, 
	adopts the \emph{Proof of Work} (PoW) 
	as a consensus mechanism, 
	in which each node in the network has to store all transactions.
	This requires high computing power and large storage space,
	and it is not realistic for IoT devices with limited cost. 
	
	\item[Latency.]
	Blockchain relies on a number of nodes to commit and confirm a transaction. 
	In general, 
	the speed of transaction generation is much slower than 
	the confirmation speed~\cite{zamaniRapidChainScalingBlockchain2018}, 
	and most transactions are in the waiting queue for confirmation,
	which increases the latency of the system as a whole.
	
	\item[Scalability.]
	Due to the latency incurred in the confirmation of a transaction,  
	the capacity of the entire network has a finite upper limit, 
	and therefore continuous addition of new nodes into 
	the network system increases congestion~\cite{karameSecurityScalabilityBitcoin2016}.
	
	\item[Transaction Cost.]
	To obtain priority in the confirmation process of transaction processing, 
	new transactions often need to pay a fee to 
	the verification node~\cite{saadMempoolOptimizationDefending2019}.
	
	\item[Interdependent Features and tradeoffs.]
	In specific IoT application scenarios, 
	while emphasizing and enhancing some DLT features,
	we often have to quantify the trade-off among features,
	such as privacy and performance. 
	In other words, 
	it is difficult to satisfy the constraints of DLT all the times. 

	\item[Cost-effectiveness and Performance.] 
	Generally speaking, 
	different DLTs have unique structures and advantageous scenarios. 
	However, 
	in practice, 
	especially for IoT enterprises, 
	there are often multiple scenarios that require DLTs. 
	The cost of using multiple DLTs simultaneously is very high, 
	and the blockchain's high latency and low throughput 
	further aggravate this situation. 
	So we need to come up with the most suitable version of DLT 
	based on the internal mechanism of IoT applications~\cite{mistryBlockchain5GenabledIoT2020}.
\end{description} 

\begin{table}[t]
	\centering
	\caption{Comparison with Existing Surveys}
	\resizebox{\textwidth}{!}{
		\begin{tabular}{p{8em}p{6em}p{7em}p{6em}p{7em}p{7em}p{15em}}
			\toprule
			Survey & \citet{atlamBlockchainInternetThings2018} & \citet{antalDistributedLedgerTechnology2021} & \citet{kannengiesserTradeoffsDistributedLedger2020} & \citet{zhuApplicationsDistributedLedger2019} & \citet{bourasDistributedLedgerTechnology2020} & \textbf{This Work} \\
			\midrule
			Research Evolution & --- & --- & --- & --- & --- & Visualize evolution paths using knowledge graphs \\
			\midrule
			Challenges & Challenges of blockchain with IoT & --- & --- & Focus on future directions rather than current challenges & Basic challenges faced by DLT in eHealth & Thoroughly identify DLT-IoT convergence challenges from each layer \\
			\midrule
			DLT for IoT & Blockchain & Blockchain\newline{}DAG\newline{}Federal ledger & --- & Blockchain\newline{}DAG & Blockchain & Novel prototypes that are IoT-friendly and quantum-safe, such \emph{IOTA} and \emph{Hashgraph} \\
			\midrule
			Industry Verticals & --- & --- & --- & Smart Home \newline{} Smart Transport \newline{} Supply Chain \newline{} Smart Health \newline{} Smart Energy & eHealth & The latest investigation of \citet{zhuApplicationsDistributedLedger2019} + Smart City, and application proposals based on DLT characteristics \\
			\midrule
			Architecture & 4-layer IoT \newline{}architecture & 3-tier DLT \newline{}architecture & Hierarchical structure & 5-layer DLT\newline{}architecture & --- & A new 6-layer architecture build on ~\cite{atlamBlockchainInternetThings2018, antalDistributedLedgerTechnology2021, kannengiesserTradeoffsDistributedLedger2020, zhuApplicationsDistributedLedger2019, bourasDistributedLedgerTechnology2020} allows for a more comprehensive projection of solutions, e.g., \emph{IOTA}. \\
			\bottomrule
		\end{tabular}%
		\label{tab:contribution}
	}
\end{table}%

The literature on DLT for IoT is quite rich~\cite{bourasDistributedLedgerTechnology2020, 
	dorriLSBLightweightScalable2019, zhuApplicationsDistributedLedger2019, 
	antalDistributedLedgerTechnology2021, kannengiesserTradeoffsDistributedLedger2020, 
	atlamBlockchainInternetThings2018}.
Most survey works~\cite{antalDistributedLedgerTechnology2021, 
	parkDAGBasedDistributedLedger2019, bourasDistributedLedgerTechnology2020, 
	kokoris-kogiasOmniLedgerSecureScaleOut2018} comprehensively reviewed the challenges 
and applications of DLT and IoT separately, 
and provided researchers with a systematic and holistic views of 
the state-of-the-art in the fields. 
In \Cref{tab:contribution}, 
we summarize existing survey works and contrast them with this work, 
on various dimensions including
DLT architecture, DLT for IoT, 
DLT characteristics, applications, etc. 
Of particular relevance to this work are the studies in
~\cite{atlamBlockchainInternetThings2018, dorriLSBLightweightScalable2019, antalDistributedLedgerTechnology2021, 
	parkDAGBasedDistributedLedger2019, bourasDistributedLedgerTechnology2020, 
	kokoris-kogiasOmniLedgerSecureScaleOut2018}. 
~\citet{dorriLSBLightweightScalable2019} 
proposed a lightweight and scalable blockchain 
optimized for the needs of the Internet of Things, 
which uses a time-based distributed consensus algorithm 
to reduce the overhead of the blockchain. 
However, 
~\cite{dorriLSBLightweightScalable2019} does not fundamentally 
address the problem of transaction cost, 
but uses advertising the revenue as a novel substitute. 
Similarly, 
\citet{parkDAGBasedDistributedLedger2019} proposed a DAG-based DLT called 
\textbf{powergraph} to reduce the delay of power transaction process in smart grid, 
and the evaluation results showed that 
\textbf{powergraph} achieves higher transaction processing speed, 
though it does not consider the case of offline transactions. 
However, 
at the same time, 
since the nodes in the network actually play the role of miners, 
the transaction fees can not be eliminated.

\citet{kokoris-kogiasOmniLedgerSecureScaleOut2018} 
proposed a sharding-based DLT called \emph{OmniLedger}, 
which optimizes the performance via parallel intra-shard transaction processing, 
without accounting for resource consumption. 
Thanks to collectively-signed state blocks, 
and low-latency ``trust-but-verify'' validation for low-value transactions,
\emph{OmniLedger}'s throughput scales linearly with increase 
in the number of active validators. 

To summarize, 
the survey works 
focusing on the hierarchical architecture of DLT and its importance for IoT 
are enlisted below:
\begin{enumerate}
	\item 
	~\citet{atlamBlockchainInternetThings2018} 
	reviewed the centralized IoT architectures, 
	compared the characteristics of blockchain and IoT, 
	with an aim to extract the benefits of implementing blockchain in IoT. 
	However, 
	a wider range of model choices, 
	such as DAG based DLT,
	were not considered.
	
	\item 
	~\citet{antalDistributedLedgerTechnology2021} 
	constructed and analyzed a 3-tier technical architecture 
	based on the decentralized application of blockchain, 
	and proposed a technical route to implement a distributed system. 
	However, 
	they did not pay attention to industry applications and 
	the challenges of DLT-IoT convergence. 
	
	\item 
	~\citet{kannengiesserTradeoffsDistributedLedger2020} 
	investigated the literature and consulted experts to 
	determine the popular DLT characteristics, 
	then comprehensively summarized the trade-offs among different characteristics. 
	However, 
	they did not consider the effects of different DLT 
	prototypes in the context of IoT. 
	
	\item 
	~\citet{zhuApplicationsDistributedLedger2019} 
	divided DLT into a five-layer architecture. 
	The difference of~\cite{zhuApplicationsDistributedLedger2019} from other studies is that 
	they distinguished blockchain with DAG-based models  
	and reviewed \emph{machine-to-machine} (M2M) economy.
	However, 
	the tight coupling of DLT characteristics with IoT applications requires further exploration and systematic survey.
	
	\item 
	~\citet{bourasDistributedLedgerTechnology2020} 
	decomposed the key players and their tasks in \emph{eHealth} in detail, 
	aiming to build an \emph{eHealth} architecture based on DLT.
	A clear architecture diagram and solution comparison are given. 
\end{enumerate}

As mentioned, 
although DAG-based IoT prototypes are emerging
~\cite{antalDistributedLedgerTechnology2021,zhuApplicationsDistributedLedger2019}, 
many recently proposed models   
have not been covered in the existing surveys.
Overall,
existing surveys have partially demonstrated existing works,
but are not comprehensive for the IoT applications of DLT, 
especially in the context of DLT-IoT convergence. 
Most importantly, 
no existing survey work discusses
the feasibility of DLT-IoT convergence scenarios and 
comprehensively envisages the development of novel DLTs for IoT applications.

\subsection{Contributions} 

Our first main contribution in this work is 
\textit{a new framework for conducting a systematic and comprehensive review
	of DLT over IoT by extending the knowledge graph approach}
\footnote{A knowledge graph is a semantic network 
	that demonstrates a set of interconnected descriptions of 
	objects, events, or concepts.}.
To better organize and understand the links of existing theories, 
proposals, gaps and research trends to date, 
we extend the knowledge graph to 
capture the structural foundation of DLT-IoT convergence by 
specifically focusing on their evolution and the current state-of-the-art. 
The outcomes enable us to develop a holistic understanding of the underlying context and 
to generate insightful visualization of the underlying evolution, 
research trends and impeding gaps for DLT-IoT convergence. 
Resulting outcomes are illustrated later (e.g., see~\Cref{KGAlgorithm} and \Cref{fig:timeline}). 

Other main contributions of this work are threefold:
\begin{enumerate}
	\item We reviewed the existing DLT prototypes. 
	In the detailed comparison, 
	we found that 
	the blockchain has inherent limitations over the IoT applications. 
	On the contrary, 
	there are several prototypes that were not fully considered to
	have great application potential in IoT.
	
	\item We comprehensively investigate the industry use cases of DLT, especially in the context of IoT. 
	Different from the existing work, 
	we extracted specific DLT characteristics 
	that needs to be prioritized for some specific industry verticals.
	
	\item We develop new insights towards 
	a comprehensive hierarchical architecture of DLT-IoT convergence, 
	which, in contrast to the existing architectures, 
	account for more detailed and integrated modules. 
	It exploits the tightly coupled relationship 
	between DLT capabilities with IoT applications. 
	
\end{enumerate}

\subsection{Organization of this paper} \label{sec-organization}

\begin{figure}[t]
	\centering
	\includegraphics[scale=0.4]{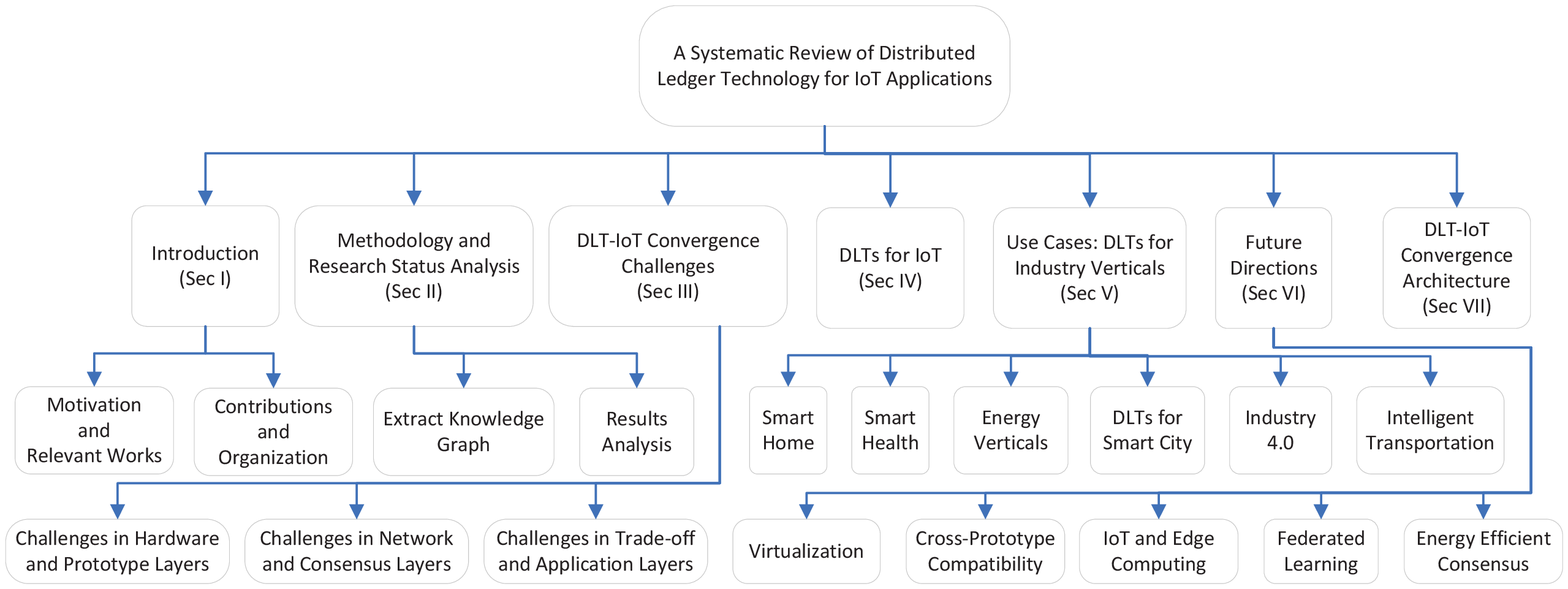}
	\caption{A high level view of the organization and structure of this paper}
	\label{fig:articlestructure}
	\vspace{-5 mm} \end{figure}

\Cref{fig:articlestructure} shows a high-level view  of the organization of this paper.
In the~\Cref{sec-architecture}, 
we develop a new six-layer DLT architecture, 
which introduces a function Layer to 
show the close coupling relationship 
between DLT characteristics and IoT applications, 
and then we discuss the challenges of DLT-IoT convergence
from the prospective of different layers in the~\Cref{sec-challenges}.
In the~\Cref{sec-dlts}, 
we introduce other DLT prototypes 
that have application potential in IoT scenarios, 
including \emph{IOTA}, \emph{Hashgraph}, 
and some blockchain variants with higher scalability, 
then discuss their differences and their respective advantages.
In~\Cref{sec-usecases}, 
we comprehensively introduce the main industries of IoT, 
such as smart home, smart city, intelligent transportation, 
smart energy, smart health, and smart manufacturing, 
then discuss the use cases and important characteristics of DLT 
in different industries.
In~\Cref{sec-discussion}, 
we investigate the recent promising conceptual models, 
such as DLT virtualization and neural ledger,
then discuss the impact of these models on IoT and future research directions.

\section{Methodology and Research Status Analysis} \label{sec-methodology}

In this section,
we analyze the literature extensively using theoretically supported techniques 
to avoid subjectivity, of which the knowledge graph is one such tool.

\citet{speelKnowledgeMappingIndustrial1999} 
developed \textit{knowledge graph} as a visualization mechanism 
that analyzes the structure of the existing domain knowledge to 
discover hidden features, unforeseen connections and gaps.
It adopts graphics to present the connections between knowledge, 
which, when used for a systematic survey of emerging technology, 
it can help researchers quickly understand the research status 
and envisage future research directions. 

Among the tools dedicated to building knowledge graph, 
\emph{Citespace} is a celebrated tool~\cite{chenCiteSpaceIIDetecting2006,chenStructureDynamicsCocitation2010}
to analyze the research connections and relationships between articles, and on that basis, 
it can generate a variety of co-citation networks 
to reveal the knowledge structure of the current field
\footnote{See {\url{http://cluster.cis.drexel.edu/~cchen/citespace/}} 
	for further information and details.}.

\emph{Citespace} connects to the Web of Science data sources, 
and we look for the keyword ``\texttt{distributed ledger}''
and designate the ``Web of Science Core Collection'', 
to search scope as ``Topic". 
We found \num{1489} articles matching the criteria.
Then,  
we export the full records and references into plain text format 
and use \emph{Citespace} to generate knowledge graphs. 

To this end, 
extending the idea proposed in~\cite{eggheTheoryPractiseGindex2006}, 
we design \Cref{KGAlgorithm} to generate a knowledge graph from the exported citations.  
It is worth noting that 
we use the modified \emph{g-index}~\cite{eggheTheoryPractiseGindex2006} to select nodes 
and then calculate the connection between the nodes by \emph{cosine} similarity 
as shown in \Cref{code: KGA: cos}. 
For the clustering analysis, 
we introduce the concept of \emph{sub-graphs} and \emph{cut function} 
without explanation in \Cref{code: KGA: subgraph}. 
In fact, 
the algorithms involved in \emph{Citespace} are far more complex than this. 
For more technical details, 
please check the early works of Chen~\cite{chenStructureDynamicsCocitation2010}.

\begin{algorithm}[h]
	\caption{Knowledge graph generation algorithm}
	\label{KGAlgorithm}
	\begin{algorithmic}[1]
		\Require Slicing citations files per year: $ \mathcal D = \{ ...,slic{e_{year}},...\}, year = \{2016,...,2021\}, k $
		\Ensure Knowledge Graph $ \mathcal G = (\mathcal V, \mathcal E) $
		\ForAll{$ slic{e_{year}} \in \mathcal D $}
		\State select nodes set $ \mathcal V $ by
		\State ${g^2} \le k\sum\limits_{i \le g} {{c_i}} ,k \in {Z^ + }$
		\State  $ \{ {A_i}\} _{i = 0}^n \leftarrow {\rm{ }}Set\ of\ papers\ that\ cites\ {c_i} $
		\EndFor
		\State Get $ \mathcal E $ by computing cosine similarity $ {w_{ij}} $ between $ A_i$ and $ A_j$
		\label {code: KGA: cos}
		\ForAll{$ c_i \in \mathcal V$}
		\ForAll{$ c_j \in \mathcal V$}
		\State $ {w_{ij}} \leftarrow {{|{A_i} \cap {A_j}|}}\big/{{\sqrt {|{A_i}| \times |{A_i}|} }} $
		\EndFor
		\EndFor
		\State Compute sum of weights of links in sub-graphs $ \mathcal G_k $
		\label {code: KGA: subgraph}
		\State $\operatorname { vol } ( \mathcal G _ { k } ) \leftarrow \sum _ { i \in \mathcal G _ { k } } \sum _ { j } w _ { i j }$
		\State Search different clusters by
		$\min_k \sum _ { k = 1 } ^ { K }  { \operatorname { cut } ( \mathcal G _ { k } , \mathcal G - \mathcal G _ { k } ) } \big/{ \operatorname { vol } ( \mathcal G _ { k } ) }$
		\State Automatic node labeling based on $frequency$
		\State Automatic cluster labeling by $Latent\ Semantic\ Indexing\ (LSI)$~\cite{deerwesterIndexingLatentSemantic1990} 
	\end{algorithmic}
\end{algorithm}

\begin{figure}[t]
	\centering
	\includegraphics[width=0.8\textwidth]{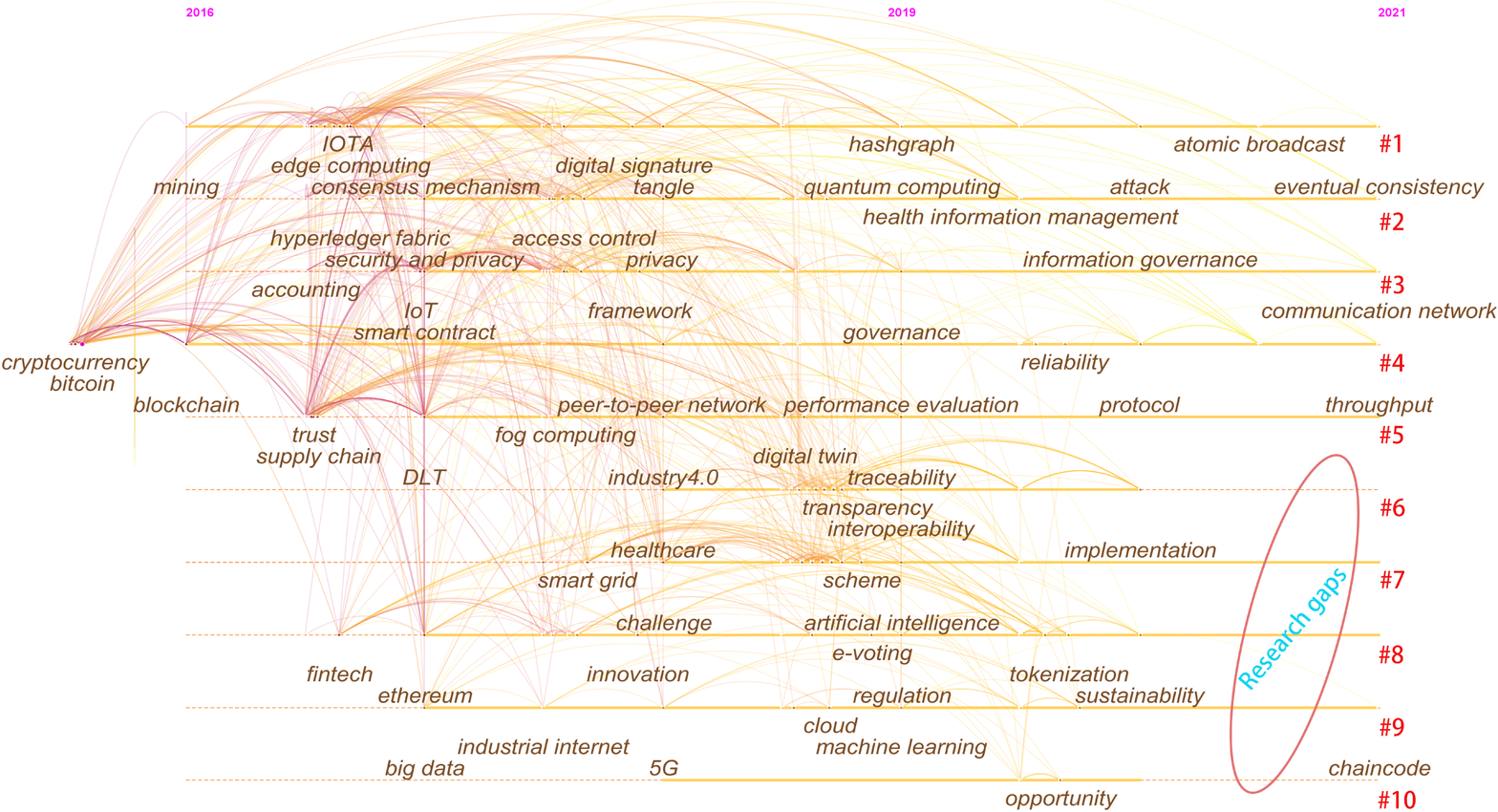}
	\caption{Knowledge graph of DLT for IoT: Timeline View}
	\label{fig:timeline}
	\vspace{-5 mm} \end{figure}

To discover and find the research trends, directions and gaps, 
we apply the timeline view and illustrate the 
generated knowledge graph using~\Cref{KGAlgorithm} in \Cref{fig:timeline}. 
From~\Cref{fig:timeline},
we can observe that 
\num{10} clusters are spreading from top to bottom. 
Specifically, 
each line represents a cluster, 
and the milestones nodes depicting important trends of research 
in each cluster are marked over the lines in chronological order.

We can clearly see in \Cref{fig:timeline} that 
the origin of everything is the cryptocurrency, 
in particular the \emph{}{Bitcoin} (BTC). 
This emergence of BTC demonstrated the ignition 
towards researchers' enthusiasm for blockchain research. 
After the real momentum in blockchain research, 
several blockchain applications with large communities and investments 
have emerged quickly, 
thus advancing DLTs into new forms such as \emph{Hyperledger}, \emph{IOTA} and \emph{Ethereum}. 

In fact,  
as the largest user community and the most mature blockchain application, 
researchers have favoured \emph{Ethereum}, 
so that it has become a cluster alone 
(see \verb!#9! in \Cref{fig:timeline}). 
However, 
due to the pressure of supervision and the growth of users, 
\emph{Ethereum} must seek breakthroughs and consensus innovations to 
meet the requirements of regulations and 
its own sustainable development, 
which is also consistent with the changes of research hotspots
shown in the knowledge graph.

The security and privacy threats faced by IoT moved research
towards adopting DLT in this field. 
However, 
hardware constraints require 
more resource friendly and scalable consensus mechanism. 
Naturally,
such a consensus mechanism shall meet the above expectations. 
As the consensus mechanism of \emph{IOTA}~\cite{popovTangle2018}, 
\emph{Tangle} has recently attracted the attention of researchers 
with its excellent performance and scalability. 
Moreover,
DLT also helps IoT to be more accountable, 
when data leakage occurs 
which helps intelligent IoT devices to perform more 
secure access control and ensure data security 
(see \verb!#1!, \verb!#2! and \verb!#3! in \Cref{fig:timeline}). 

Another exciting observation in \Cref{fig:timeline} is 
the emergence of new technologies, 
such as 5G, big data, machine learning, cloud, and artificial intelligence. 
They have brought new opportunities for DLT over IoT 
(see \verb!#8! and \verb!#10! in \Cref{fig:timeline}).
In addition, 
quantum computing brings opportunities with more severe security challenges 
to the existing consensus mechanisms and encryption algorithms. 
We can clearly see that 
there are few nodes in the lower right corner of the knowledge graph, 
which demonstrates research gaps from 2020 to 2021. 
To this end,
we provide a comprehensive summary of the field's hierarchy, 
from features to applications, 
and discuss challenges and opportunities to deliver material support and 
sources of inspiration for new research.

\section{DLT- IoT Convergence Challenges} \label{sec-challenges}

The DLT-IoT convergence faces many challenges.
These challenges seem to be chaotic, 
but in fact, 
we can identify their sources, 
which will help us propose targeted improvements for each challenge. 
We will also discuss the challenges of DLT-IoT convergence 
later in \Cref{sec-architecture}. 
In this section, 
we will identify and distinguish the corresponding challenges 
based on the proposed DLT for IoT architecture, 
as shown in \Cref{tab:challenges}.

\begin{table}[htbp]
	\centering
	\caption{Challenges of DLT-IoT convergence in different layers}
	\begin{tabular}{lll}
		\toprule
		\textbf{Hardware \& Prototype} & 
		\textbf{Network \& Consensus } & 
		\textbf{Trade-off \& Application}  \\
		\midrule
		Resource limitation~\cite{bartolettiGeneralFrameworkBlockchain2017} & Scalability~\cite{brewerRobustDistributedSystems2000,gobelIncreasedBlockSize2017} & 
		CH trade-off~\cite{kannengiesserTradeoffsDistributedLedger2020} \\
		Energy consumption~\cite{devriesBitcoinGrowingEnergy2018} & 
		Transaction fee~\cite{saadMempoolOptimizationDefending2019} & 
		Coupling of CH and App~\cite{kannengiesserTradeoffsDistributedLedger2020} \\
		Generality of prototype~\cite{mistryBlockchain5GenabledIoT2020} & 
		Throughput~\cite{brewerRobustDistributedSystems2000} &  \\
		& Lantency~\cite{gobelIncreasedBlockSize2017} &  \\
		\bottomrule
	\end{tabular}%
	\label{tab:challenges}%
\end{table}%

\subsection{Challenges in Hardware and Prototype Layers} 
\label{subsec-challenginhardware}

In the prototype layer, 
we have introduced several attractive designs of DLT with different structures. 
But when we examine hardware and prototype layers 
from more stereoscopic perspectives, 
we have found that there is a subtle relationship between them, 
which gives rise to some application challenges, 
as shown in \Cref{tab:challenges},
including resource limitation, 
energy consumption, 
and generality of the prototype. 
Therefore, 
we will reveal these challenges more specifically.

When applying DLT in IoT, 
we must consider resource limitations. 
According to the estimates in~\cite{bartolettiGeneralFrameworkBlockchain2017}, 
the size of the complete BTC transaction history 
in the past ten years has exceeded 300GB.
At the same time,
the total data size of \emph{Ethereum} has exceeded 5TB, 
and the size of each node has also reached 165GB
\cite{mutarImplementationNationalCryptocurrency2019}. 
To make matters worse, 
as new nodes join and transactions continue, 
the size of the blockchain ledger will become larger and larger. 
This is unacceptable for IoT devices with limited storage space.
The usual solution is to distinguish the nodes functionally. 
For example, 
in the blockchain, 
only full nodes store the complete chain. 
\emph{IOTA} introduces the concept of local snapshots. 
\emph{Permanodes} will store all transaction data, 
while Full nodes will only store transaction data 
since the last network snapshot.
However, 
this method will cause some nodes to rely on other nodes to 
inform the status of the current distributed ledger, 
which will lead to the centralization of the network. 
In addition, 
too large a chain will also lead to the growth of synchronization time, 
especially when IoT devices are usually narrow bandwidth.

Energy consumption is another concern. 
The PoW consensus mechanism is adopted in the blockchain, 
which means that a transaction needs much computation 
by CPU or GPU before it is confirmed. 
As we all know, 
the power of CPU and GPU is quite high. 
Maintaining the continuous high load operation of 
these devices will 
bring huge energy consumption~\cite{devriesBitcoinGrowingEnergy2018}. 
IoT devices with weak computing power and limited power 
are difficult to be competent for such tasks. 
On the other hand, 
it will also bring environmental protection problems~\cite{trubyDecarbonizingBitcoinLaw2018}.
Of course, 
we can use other consensus mechanisms instead of PoW to alleviate this situation. 
For example,
\textrm{Ethereum} 2.0 will use PoS instead of PoW.
However, 
the mainstream blockchain consensus mechanism appears to hold PoW for some time now.

\subsection{Challenges in Network and Consensus Layers} 
\label{subsec-challenginnetwork}

For a long time in the past, 
the consensus mechanism based on PoW has ensured the security of DLT. 
However, 
with the vigorous development of DLT, 
the number of users has also ushered in explosive growth, 
and the incredible number of transactions brought by it has put forward more 
stringent requirements for the DLT network. 
It seems that 
the current DLT network infrastructure and network topology are 
facing more and more obvious challenges. 
As shown in \Cref{tab:challenges}, 
Therefore, 
we try to explain from the architecture why DLT is facing these challenges, 
such as latency, transaction fee, throughput, scalability, 
as we have shown in \Cref{tab:challenges}.

In these two layers, 
we must realize that the network bandwidth between nodes is not always high, 
especially for IoT.
Imagine in a DLT network, 
assuming that it takes \num{2} minutes for a new transaction to be accepted by all nodes, 
and a new transaction will be generated every minute in the network, 
then there will be such a situation: 
a node has already recorded a new transaction 
when it has not received the last transaction data. 
In the blockchain, 
since the blocks are added one by one in chronological order, 
this situation will lead to the fork of the blockchain, 
which reduces the security of the blockchain, 
because the computing power of the entire network will be apportioned. 
The attack can be completed with a computing power lower than \SI{51}{\percent}.
Therefore, 
in order to ensure security, 
the blockchain sets the block generation speed of the entire network 
at a deficient number, 
which means that the ability of the DLT system to process transactions, 
that is, throughput is very low, 
and accordingly, the difficulty of PoW will be severe. 
However, 
not all nodes have the ability to perform such complex calculations, 
so the verification of transactions will be completed 
by nodes with high computing power, 
but this brings other problems.

When only part of the nodes verify transactions, 
it naturally means that the speed of transaction verification maybe 
slower than the speed of transaction generation. 
In this case, 
many new transactions will be queued for verification, 
and the latency of the entire DLT will be quite high.
With the addition of new nodes, 
the number of new transactions will increase, 
which will cause network congestion and even system paralysis.
Therefore, 
for BTC that uses the PoW consensus mechanism, 
the scalability is quite poor.
At the same time, 
in order to obtain the priority of transaction confirmation, 
the node will have to pay a fee to the verification node, 
which increases the cost of DLT.

The DLT prototype based on DAG claims to eliminate transaction fees, 
because the DAG network topology makes transaction processing parallel, 
so there is no need to perform complex PoW calculations, 
and each node will participate in the transaction confirmation process. 
With the extension and entanglement of the DAG network, 
the effect of collaborative verification becomes more and more obvious, 
so it has extremely high throughput and scalability while maintaining low latency. 

\subsection{Challenges at the Function and Application Layers} 
\label{subsec-challenginapplication}

When we review the unlimited potential of the combination of DLT and IoT, 
we can see potential for many tightly coupled IoT application scenarios. 
However, 
this relationship and the fragmentation characteristics of DLT bring 
challenges to the application, as shown in \Cref{tab:challenges}.

\begin{figure}[t]
	\centering
	\includegraphics[scale=0.357]{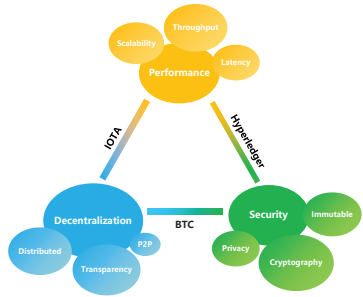}
	\caption{Trade-off relationship between DLT Characteristics}
	\label{fig:trade-off}
	\vspace{-5 mm} \end{figure}

Using \Cref{fig:trade-off}, 
we illustrate and explain this complex interrelationship 
and construct a Ternary chart to show 
the trade-offs on DLT features. 
Although the combination of DLT and IoT brings encouraging benefits, 
the hierarchical relationship of architecture brings 
the dependence on the native characteristics of DLT. 
Furthermore, 
limited resources make it impossible for us to achieve 
all DLT characteristics optimally in the application. 
On the contrary, 
improving one or more of them has to weaken the others, 
which leads to a trade-off.
For example, 
as mentioned above, 
in order to ensure the security of the blockchain, 
the continuous addition of new nodes will lead to the increasing difficulty of PoW, 
resulting in network congestion and poor performance 
~\cite{brewerRobustDistributedSystems2000,gobelIncreasedBlockSize2017}. 
Because of the trade-off, 
DLT applications are strongly coupled to DLT characteristics. 
In other words, 
applications will make trade-offs on DLT characteristics 
according to the requirements of scenarios, 
to determine their own architecture.
For example, 
for high scalability and no transaction fee, 
DLT prototypes based on DAG can be used. 
\Cref{fig:trade-off} intuitively describes this trade-off and coupling relationship. 
As summarized in \Cref{tab:dltcharacter}, 
these three types of characteristics can be regarded as the vertices of a triangle, 
and in the application, 
we can only take a certain edge of the triangle, 
which means that DLT can improve at most two of them, 
and it does not perform well on the other type
~\cite{kannengiesserWhatDoesNot2019,zupanHyperpubsubDecentralizedPermissioned2017}. 
For example, 
\emph{IOTA} has chosen better decentralization and performance, 
so it is not as secure as BTC and Hyperledger. 
Hyperledger has chosen better performance and security, 
but it is more centralized to a certain extent, 
and BTC has chosen better Decentralization and security, 
but the difficult PoW brings poor performance.

\section{Modern DLTs for IoT} \label{sec-dlts}

Generally speaking, 
the blockchain uses PoW 
as the underlying consensus mechanism~\cite{wrightBitcoinPeertoPeerElectronic2008}. 
In most blockchain application scenarios 
(such as BTC), 
miners consume a large amount of computing power 
and simply use it for competition, 
so this leads to a waste of computing resources 
in the distributed network~\cite{luuPowerSplittingGames2015}.  
To alleviate this problem, 
some improved consensus mechanisms 
have designed an incentive system 
different from the blockchain-consuming resources 
within the network, 
such as \emph{Proof of Stake} (PoS) 
~\cite{kingPpcoinPeertopeerCryptocurrency2012}, 
\emph{Proof of Elapsed Time} (PoET), 
\emph{Practical Byzantine fault tolerance} (PBFT)~\cite{castroPracticalByzantineFault2002}, 
\emph{Federated Byzantine Fault Tolerance} (FBFT), 
it is worth noting that 
PDFT is considered to be used in IoT, 
but unfortunately, 
it still does not scale. 
The high demand for computing resources and scalability 
makes it difficult for blockchain to 
be implemented on a large scale, 
especially for the application of 
~\cite{laoSurveyIoTApplications2020} in IoT. 
Therefore, 
some researchers have begun to explore that 
there is no need for miners to participate, 
no transaction fees, and good scalability. 
Most of these systems are based on \emph{directed acyclic graphs} (DAG). 
Unlike the chain structure formed 
by connecting blocks in chronological order of the blockchain, 
DAG is a network structure in which 
transactions are connected. 
~\cite{churyumovByteballDecentralizedSystem2016} first introduced DAG into DLT. 
After that, 
many DAG-based or DLT systems similar to DAG, 
such as \textbf{IOTA}~\cite{popovTangle2018}, 
\textbf{HashGraph}~\cite{bairdSwirldsHashgraphConsensus2016}, 
\textbf{NANO}~\cite{lemahieuRaiBlocksFeelessDistributed2017} etc. 

\subsection{IOTA} \label{subsec:IOTA}

\begin{figure}[htbp]
	\centering
	\subfloat[The original Tangle network]{
		\includegraphics[width=0.4507\textwidth]{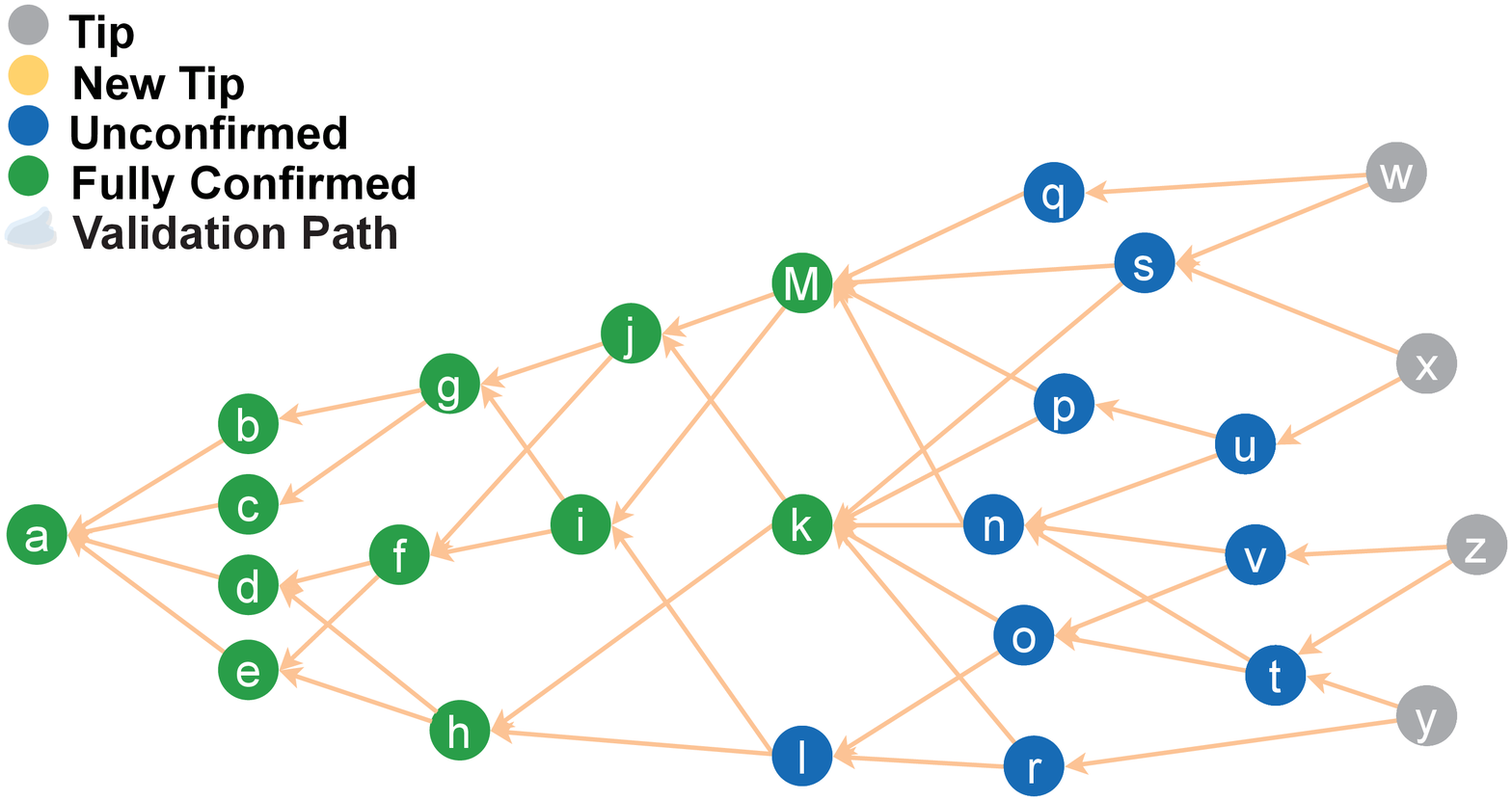}
		\label{fig:tangle1}
	}
	
	\subfloat[Tangle network after adding two new tips]{
		\includegraphics[width=0.4507\textwidth]{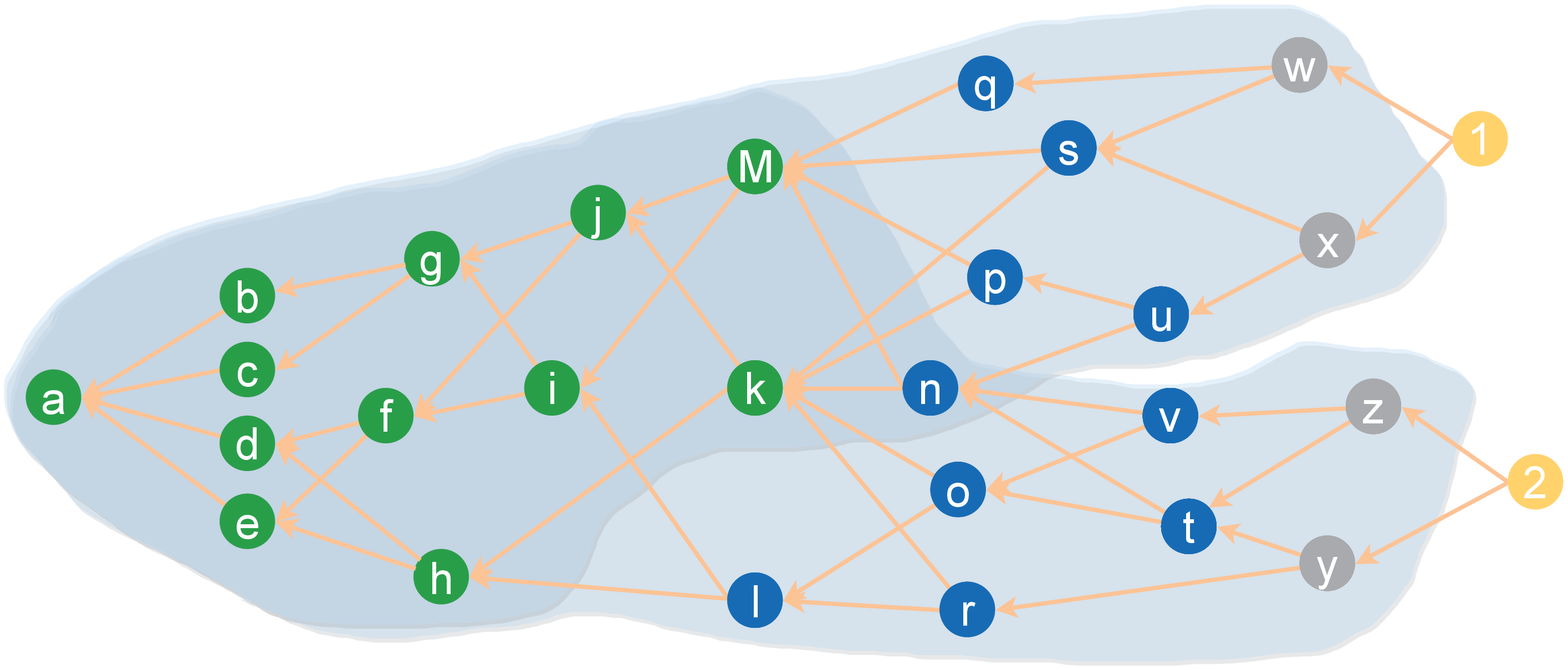}
		\label{fig:tangle2}
	}
	\caption{Tangle's consensus process}
	\label{fig:Tangle's consensus process}
	\vspace{-5 mm} \end{figure}

Among all DAG-based DLTs, 
\emph{IOTA} is considered the most promising~\cite{antalDistributedLedgerTechnology2021}, 
because it is specially designed for IoT and has great application prospects in the industry 4.0 era. 
\emph{IOTA} uses DAG and a well-designed consensus mechanism. 
These two form a distributed ledger that becomes the \emph{Tangle}, 
which eliminates the part of the blockchain that relies on miners for mining. 
Every node in the tangle network is assumed to 
``hope that their transactions will be confirmed as soon as possible'', 
which drives them to participate in the verification and maintain network security. 
In the \emph{IOTA} transaction process, 
when a node tries to add a transaction to the tangle network, 
it needs to find two transactions that other nodes have not confirmed, 
verify its validity through a small amount of calculation, 
and then point its transaction to these two transactions. 
At this time, 
the transaction will be added to the network, 
waiting for subsequent transactions to be verified. 
This method is officially called ``\emph{Tip Selection Algorithm}'' by 
\emph{IOTA}~\cite{kusmierzPropertiesTangleUniform2019}, 
and \Cref{fig:Tangle's consensus process} shows the 
network structure and consensus process of \emph{Tangle}.

In the tangle network shown in \Cref{fig:tangle1}, 
the transactions marked in green have been fully confirmed, 
the transactions marked in blue are in a partially confirmed state, 
and the tips marked in grey represent unconfirmed transactions in the \emph{Tangle}. 
In \Cref{fig:tangle2}, 
there are two newly generated tips, 
denoted by 1 and 2, respectively. 
As mentioned before, 
if these two tips want to be added to the tangle network, 
they must first verify two unconfirmed transactions. 
Tip 1 selects transaction ``W" and transaction ``X", 
and then verifies the validity of them and their verified transactions 
(Q, S, U, P, M, K, etc.) through a few calculations, 
so it sets itself Point to ``W" and ``X," 
the light blue area indicates the verification path of tip 1. 
And tip 2 chose transaction ``Z" and transaction ``Y" and did the same thing. 
Since the transaction ``N" has been verified by all current tips, 
it becomes a fully confirmed state. 
On the one hand, 
the nodes in the tangle network do not need to verify all transactions, 
and the verification can be parallel 
(for example, 
the verification process of tip 1 and tip 2 can be performed at the same time), 
so the \emph{Tangle} can achieve faster verification speed and more High throughput, 
on the other hand, 
as shown in \Cref{fig:tangle2}, 
the verification paths of new tips will overlap and cross each other, 
so transactions will be repeatedly confirmed, 
and collaborative verification of the entire network will appear.

\subsection{Hashgraph} \label{subsec:Hashgraph}

\emph{Hashgraph} also uses DAG as the network structure, 
which is the same as \emph{IOTA}, 
but its most special point is that it does not require any authority nodes. 
This is often people's worry about DAG-based DLT prototypes, 
For example, 
the validator node in \emph{IOTA} and the witness node in byteball.
At the same time, 
the consensus process of \emph{Hashgraph} is 
essentially an asynchronous BFT algorithm (ABFT), 
so \emph{Hashgraph} can be as secure as a blockchain,
which has been proven mathematically~\cite{bairdHederaGoverningCouncil2018}. 
It is well known that PBFT has poor scalability 
because its communication complexity is $ o({n^2}) $. 
When the number of nodes in the network increases, 
the messages that need to be synchronized will increase exponentially, 
which will cause serious network congestion. 
Because the consensus process of \emph{Hashgraph} is completely asynchronous 
and based on the DAG network structure, 
it achieves higher scalability and throughput. 
The official data shows that the throughput of \emph{Hashgraph} 
can reach an astonishing 250K TPS in a real environment. 
~\footnote{\url{https://hedera.com/}}

The ability to maintain such high security while achieving extremely 
fast transaction processing relies on two main mechanisms of the \emph{Hashgraph}, 
namely \emph{Gossip about Gossip} and \emph{Virtual Vote}~\cite{bairdSwirldsHashgraphConsensus2016}. 
\emph{Gossip} means that information spreads by each member 
repeatedly choosing another member at random, 
and telling them all they know. 
Let us simply assume that there are three nodes in the network:
node one sends all the information it has to node two through the gossip protocol, 
and node two also sends all the information it has to node three, 
which contains the information received from node one, 
so it is called \emph{gossip about gossip} (GaG). 
After GaG, 
all nodes in the \emph{Hashgraph} network are full nodes, 
so each node can independently perform virtual voting, 
and finally will get the same result, 
that is, a consensus has been reached. 

\begin{figure}[htbp]
	\centering
	\includegraphics[scale=0.4]{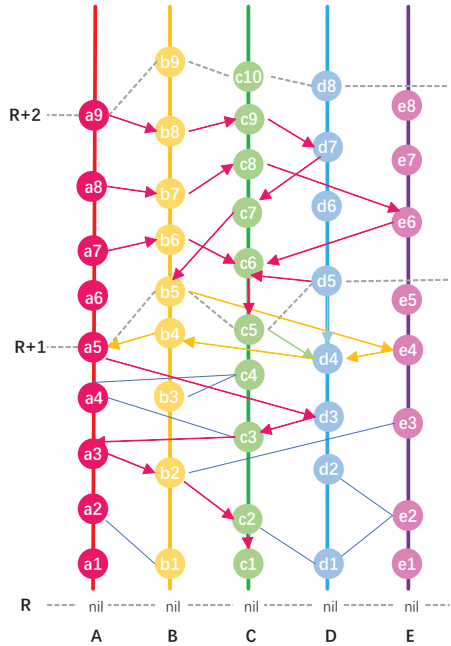}
	\caption{The consensus process of \emph{Hashgraph}: At first, 
		a1, b1, c1, d1, and e1 are the 	witnesses of the round $R$ because they 
		are the first event of each node in round $R$. 
		After that, 
		The a5 can see more than $\frac{2}{3}$ of the witnesses, 
		so enter round $R + 1$, and b5, c5 and d5 are the same. 
		At the same time, 
		a5, b5, c5, and d5 strongly see c1, 
		so c1 becomes a famous witness. 
		Finally, 
		a9 strongly sees a5, b5, c5 and d5, 
		so c1 as a famous witness can be confirmed immediately.}
	\label{fig:hashgraph}
	\vspace{-5 mm} \end{figure}

\emph{Hashgraph} uses a concept similar to block, called event. 
An event contains a timestamp, a transaction set, its own hash value, 
and the hash value of an event received from another node.
We can create timelines for all nodes separately, 
and events are added to the timeline from bottom 
to top in the order of generation. 
At the same time, 
due to GaG, 
events of different nodes will be continuously connected, 
thus forming a \emph{Hashgraph}, 
as shown in \Cref{fig:hashgraph}.
Due to the connection relationship, 
if a certain sub-event m can be traced forward to the ancestor event n, 
then it is called m seeing n. 
In particular, 
if the path of forward tracing spans more than $\frac{2}{3}$ of the nodes in the network, 
then it is called m strongly seeing n, 
and events generated by a certain node are evident 
to subsequent events of that node.

To complete virtual voting, 
\emph{Hashgraph} divides events into different rounds according to their visible state.
When an event is created, 
it is in the same round as the parent event, denoted as round $R$, 
and the first event created in a certain round is called a witness, 
If an event strongly sees more than $\frac{2}{3}$  of the witness events, 
this event immediately enters the next round, denoted as round $R+1$.
Furthermore, 
if a witness event in the round $R$ can be strongly seen by 
more than $\frac{2}{3}$ of events in the round $R+1$,
then it becomes a famous witness event.
In fact, 
this is a voting process.
The same process will continue in the round $R+2$. 
Since each node saves a copy of \emph{Hashgraph}, 
they can independently perform the above virtual voting process locally, 
and as mentioned above, 
it has been proven that 
they will eventually get the same voting 
result~\cite{bairdSwirldsHashgraphConsensus2016}.

\subsection{Other DLT advancements} \label{subsec:blockchain}

We consider the ongoing advances in DLT 
prototypes and their practicability over 
different IoT scenarios considering enhancements 
such as \emph{sharding}, \emph{edge computing}, \emph{fog computing}, etc.
As mentioned earlier, 
blockchain has some inherent disadvantages over IoT applications, 
such as insufficient scalability and 
low throughput~\cite{zhengBlockchainChallengesOpportunities2018}. 
Therefore, 
it is often difficult to apply it directly over the IoT system. 
At the same time, 
blockchain obviously needs to be further improved either because 
its scale has reached the design 
bottleneck~\cite{alqahtaniBottlenecksBlockchainConsensus2021} or 
for the practical needs of integration with IoT.
In fact, 
the current expansion plan of the blockchain mainly 
includes \emph{sharding}~\cite{luuSecureShardingProtocol2016} and 
\emph{off-chain}~\cite{poonBitcoinLightningNetwork2016}.

\emph{Sharding} was originally a concept in 
databases~\cite{corbettSpannerGoogleGlobally2013}, 
which refers to dividing the database into multiple parts 
of the same mode and different content, 
thereby distributing the load and speeding up query time~\cite{baguiDatabaseShardingProvide2015}.
Inspired by database fragmentation, 
people put forward such an idea that the nodes and transactions 
in the blockchain network are divided into different sharding networks, 
so that the parallel verification of transactions and the 
linear growth of the scalability of the blockchain can be realized
~\cite{kokoris-kogiasOmniLedgerSecureScaleOut2018}, 
which is called network sharding and transaction sharding. 
Although network sharding and transaction sharding have increased 
the throughput of the blockchain to a certain extent, 
they have not solved the resource limitation problem. 
The increasing size of the blockchain has brought a heavy burden on storage. 
In this context, 
state sharding was proposed, that is, 
while network sharding and transaction sharding are performed, 
the storage state sharding is also performed, 
which means that each node only stores the data of the \emph{sharding} it belongs to, 
rather than the data of complete blockchain, 
so that the data growth will be evenly spread across all shardings.

\begin{figure}[htbp]
	\centering
	\includegraphics[width=0.65\textwidth]{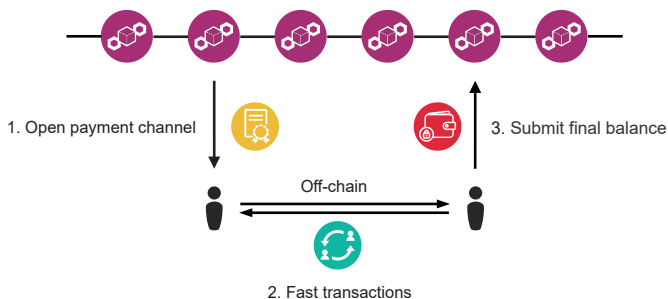}
	\caption{Lightning Network: Off-chain two-way payment channel}
	\label{fig:ln}
	\vspace{-5 mm} \end{figure}

Off-chain uses a different idea, 
that is, 
to reduce the number of transactions that the blockchain needs to process. 
Therefore, 
Off-chain requires both parties to 
handle a series of transactions on their own, 
while the blockchain only stores the final state of this series of transactions. 
Off-chain has two schemes: side chain and state channel. 
In the side chain, the user and the maintainer are not the same, 
but in the state channel, the user and the maintainer are the same. 
BTC built a second-layer payment protocol based on state channels, 
which is called \emph{Lightning Network}.  
~\footnote{\url{https://lightning.network/}}
As shown in \Cref{fig:ln}, 
in the Lightning network, 
both parties to a transaction can open a state channel dedicated to two-way payments, 
and quickly complete multiple transactions through smart contracts, 
and blockchain only verifies and stores the final state, 
thus greatly reducing the traffic 
on the BTC network~\cite{poonBitcoinLightningNetwork2016}.
The side chain refers to other blockchains that comply with the side chain protocol, 
which allows BTC to transfer from the main chain to other blockchains safely 
and to return to the main chain in the future.

Overall, 
we need to compare the above-mentioned DLTs from several aspects. 
Such a comparison can draw important conclusions, 
such as 
\textit{Which DLT prototype is more promising for which IoT applications?} 
and 
\textit{What are the necessary characteristics of DLT 
	that makes it more suitable for the IoT network consideration?}, 
among others.

The massive device scenarios of IoT require DLT to have high performance, 
including high scalability, high throughput, low latency, and low resource consumption. 
However, 
blockchains, such as BTC, 
often perform poorly in these aspects due to their chain structure. 
Therefore, 
some people choose a completely different structure from the blockchain, 
such as the DLT prototype based on DAG. 
At the same time, 
others choose to improve some links of the blockchain, 
so they propose some off-chain methods to reduce or even out the number of 
transactions to be processed, 
such as \emph{lightning network}, \emph{side chain}, and \emph{sharding}.

At present, 
the DLT prototype based on DAG is completely different 
from the blockchain in the network structure, 
and theoretically has the best performance, 
while eliminating transaction fees, 
so it is suitable for payment scenarios under the IoT.
Among them, 
the most well-known are \emph{IOTA} and \emph{Hashgraph}. 
\emph{IOTA} uses a consensus mechanism called \emph{Tangle}. 
With the extension of the network and the increase of nodes, 
the speed of transaction verification will become faster and faster. 
At that time, 
transaction fees will be completely eliminated. 
\emph{Hashgraph} uses a consensus mechanism called \emph{Gossip about Gossip}, 
and a virtual voting mechanism that can be performed locally 
realizes a completely asynchronous Byzantine fault tolerance with high security. 
In addition, 
\emph{IOTA} is open source and has a large developer community, 
while \emph{Hashgraph} is a commercial solution with a registered trademark. 

\emph{Ethereum} 2.0 will use the PoS consensus mechanism 
and \emph{sharding} to improve scalability. 
However, 
\emph{sharding} may bring some potential security concerns. 
In the case of state sharding, 
communication complexity may even offset the benefits. 
BTC uses a lightning network and side-chain to solve the scalability problem. 
However, 
neither \emph{Ethereum} nor blockchain has really implemented 
these solutions on the main chain. 

\section{Use Cases: DLTs for IoT Industry Verticals} \label{sec-usecases}

\begin{figure}[htbp]
	\centering
	\includegraphics[width=0.7\textwidth]{"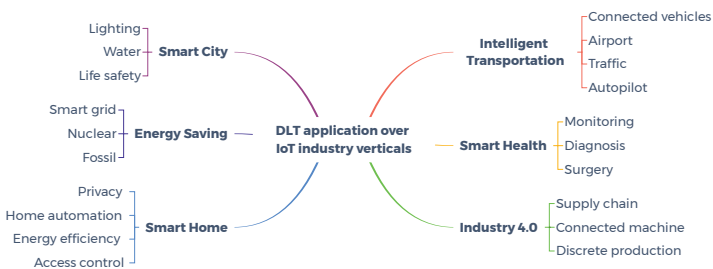"}
	\caption{DLT application over IoT industry verticals}
	\label{fig:iotapplication}
	\vspace{-5 mm} \end{figure}

With the growing emergence of mechanization and electronic information technology, 
increased productivity and the evolution of the fourth industrial revolution 
have already been evident~\cite{financeIndustryChallengesSolutions2015},
In contrast to the previous industrial revolution,  
the primary goal of Industry 4.0 is not to replace the existing manufacturing technologies 
but to utilize the advancements in information and communication technologies 
(such as 5G and IoT) to establish  high speed and low delay interconnections 
in networks by using existing assets or technologies~\cite{trappeyIoTPatentRoadmap2017}. 
This reforms traditional industries more and enhances their intelligence. 
As shown in \Cref{fig:iotapplication}, 
we sorted out the main application opportunities of IoT 
and discussed the application progress of DLT.

\subsection{Smart Home} \label{subsec:dlthome}

All kinds of smart devices in smart home 
(such as smart door lock, smart camera, smart gateway, etc.) 
are connected to each other through the Internet of things, 
which can automatically respond to people's commands or actions, 
so it can provide a more comfortable and 
safe way of life~\cite{komninosSurveySmartGrid2014}. 
In our literature review of smart homes using knowledge graph, 
the following characteristics are frequently mentioned.

\begin{description}
	\item[Privacy] 
	Ensure that personal data stored in IoT smart devices will not be tampered with, 
	and private data will not be leaked without the permission of the data owner.
	
	\item[Access control] Including identity management and identity verification,
	that is, providing different data access permissions for different identities
	and secure, anonymous access.
\end{description}

In the existing research, 
many systems are based on blockchain, 
but blockchain requires high computing resources and large storage space, 
which are limited on IoT devices, 
so some studies have proposed improved systems and evaluated their performance. 
For example, 
~\citet{dorriBlockchainIoTSecurity2017} divided the DLT application 
in the smart home situation into three main layers, 
studied the core components and functions of each layer in-depth, 
and proposed a lightweight blockchain system, 
and through simulation, 
it is verified that the system has higher security and privacy. 
~\citet{aungReviewEthereumSmart2017} made some changes based on the research of 
~\citet{dorriBlockchainIoTSecurity2017}. 
They used Ethereum to build a blockchain, 
used a private chain instead of a public chain, 
local storage instead of cloud storage, 
and added a smart home miner, 
the purpose is to store strategies for data flow or transaction management, 
but they did not evaluate the security of the system. 
~\citet{quHypergraphBasedBlockchainModel2018} takes 
into account the storage limitations of IoT devices, 
abstracts the blockchain structure from the hypergraph, 
treats each home as a node in the blockchain and uses hyper-edge storage, 
thereby reducing the entire network. 
The data storage is converted to part of the network storage, 
which reduces the storage consumption of IoT devices and also solves some security issues. 

In general, 
smart home systems based on blockchain use smart contracts and access tokens 
to complete access control, 
but they are immutable and accessible. 
Attackers can perform attacks by analyzing their execution methods, 
so some research suggests safer methods. 
For example, 
~\citet{samaniegoDetectingSuspiciousTransactions2019} uses metadata and event 
handlers instead of smart contracts and chain codes to reduce the overhead of the blockchain. 
~\citet{linHomeChainBlockchainBasedSecure2020} proposed a blockchain 
system based on group signature and message authentication code, 
which can perform anonymous identity verification on group members 
and identity verification on home gateways, 
and prove the system. 
With anonymity, 
traceability and confidentiality, 
it meets the requirements of smart homes for security and privacy. 
Other studies have proposed a DAG-based distributed ledger system for 
smart homes to solve the scalability problem. 
For example, 
~\citet{fanScalableDAGbasedDistributed2019} 
followed IOTA's reference implementation and 
used \emph{Tangle} to build an in-house transaction network 
for processing distributed energy (DERs) transactions. 

\subsection{Smart Health} \label{subsec:dlthealth}

The medical and health industry generates and stores numerous 
medical records or personal health data from patients. 
These data are susceptible and are the personal privacy of patients and need to be avoided. 
However, 
in practice, 
these data are often not controlled by the patients themselves, 
but managed and held by medical institutions. 
This causes the separation of ownership and management rights, 
which leads to a series of privacy risks, 
such as tampering with health records and privacy leakage. 
On the other hand, 
patients are not only seen in one medical institution, 
so there is a real need for sharing health data. 
However, 
under the traditional centralized model, 
it is difficult for medical institutions to share safe and reliable data. 
In this case, 
distributed ledger technology is considered to be a technology 
that is potentially revolutionary 
for the entire healthcare system 
by safely distributing health record management~\cite{mettlerBlockchainTechnologyHealthcare2016}. 
In our literature review of smart Health, 
the following characteristics are frequently mentioned.

\begin{description}
	\item[Decentralization] The patient becomes a node, 
	and the health data is owned and controlled by the patient himself.
	
	\item[Privacy] Health data cannot be accessed or illegally modified 
	without the patient's consent. 
	but can be shared safely and conveniently with the patient's permission.
\end{description}

In the centralized electronic health record management system, 
there are some privileged administrators or user accounts. 
The owners of these accounts may intentionally or maliciously abuse their rights, 
leading to the leakage of patients' private data. 
The report shows that the average proportion of data breaches involving internal personnel 
in medical institutions even reaches \SI{56}{\percent}~\cite{widupVerizonDataBreach2018,widup2018VerizonProtected2018}. 
In order to alleviate these problems, 
many researchers have proposed a decentralized and access control-based DLT system. 
~\citet{morelliAuditBasedAccessControl2019} implemented a DLT prototype 
based on auditing and Hyperledger Fabric to perform access control. 
Ideally, 
the following three access control enforcement mechanisms can be met 
to identify malicious actions by privileged users: 
\begin{inparaenum}[i)]
	\item Decentralized policy definition and management;
	\item Distributed policy evaluation;
	\item Enforce and log integrity audit. 
\end{inparaenum}

\citet{leemingLedgerMePersonalizing2019} proposed a reference architecture 
called ``\emph{My Ledger}'', 
which is based on the blockchain and is divided into a data layer and a blockchain layer. 
It is worth noting that in the blockchain layer, 
The patient grants or revokes access to data through a specific APP, 
and the hash record generated by the interactive operation 
will be stored in the blockchain for possible verification processes. 
At the same time, 
in general, 
these details are not public.

\citet{bourasDistributedLedgerTechnology2020} comprehensively reviewed 
the development of existing centralized identity management and 
distributed identity management models, 
and discussed the application potential and 
latest developments of DLT in electronic health (\emph{eHealth}), 
and then two main distributed identity models are given: 
the general structure of Self-Sovereign Identity and Decentralized Trusted Identity. 
Simply put, 
Self-Sovereign Identity is a user-centric distributed model that allows users to 
fully control their own identity and information data. 
Users, claim issuers, and relying parties 
are three main participants in this system. 
A user can obtain a distributed identifier (DID) at any time, 
and any other participant who wants to access user identity information 
must first go through user authorization~\cite{muhleSurveyEssentialComponents2018}.

\subsection{Energy Management} \label{subsubsec:dltenergy}

With the advancement of smart grid, 
renewable energy and intelligent response load equipment technology, 
the energy industry is also facing tremendous changes. 
Individual producers have participated in the energy network, 
and the energy network has changed from a central to a distributed one. 
On this basis, 
the energy system is composed of producers, 
consumers and distributed energy resources need a credible market for 
energy transactions~\cite{ziaMicrogridTransactiveEnergy2019}. 
Our knowledge graph recommends the following characteristics.
\begin{description}
	\item[Trust] The electricity consumers in the market are always confident 
	that they will not be deceived.
	
	\item[Privacy] Making a transaction will never lead to the disclosure of your privacy.
	
	\item[Performance] The system can handle large-scale transactions quickly and reliably.
\end{description}

~\citet{popBlockchainBasedDecentralized2018} used \emph{Ethereum} 
to build a blockchain prototype, 
which collects and stores energy consumption information from IoT smart metering devices, 
and defines rules that can automatically execute rewards or penalties 
through smart contracts, 
and balance energy demand and energy supply. 
Experimental results show that 
the prototype can track demand response signals with high precision, 
while reducing the energy flexibility required for convergence. 
However, 
blockchain-based microgrid energy transactions face privacy concerns, 
so~\citet{hassanDEALDifferentiallyPrivate2020} used differential privacy 
technology to ensure that 
no private information of any transaction participant can be easily inferred, 
and proposed a blockchain-based microgrid energy auction system called DEAL, 
which maintains the overall network benefits and social welfare. 

In order to solve the shortcomings of high latency and 
low performance in the blockchain structure, 
some scholars use DAG to design smart energy systems. 
~\citet{parkBlockFreeDistributedLedger2019} discussed 
the possibility of DAG being applied 
in P2P energy trading platforms from the perspective of \emph{IOTA} use cases. 
In further work, 
Park and Kim~\cite{parkDAGBasedDistributedLedger2019} proposed a DAG-based smart grid. 
The distributed ledger is called \emph{PowerGraph}. 
Each transaction in \emph{PowerGraph} is individually confirmed 
through a well-designed consensus algorithm. 

Another emerging model is the \emph{vehicle-to-grid} network (V2G)~\cite{kemptonElectricVehiclesNew1997}. 
In the V2G network, 
electric vehicles use two-way charging and purchase energy from the grid 
or sell energy to the grid according to 
the situation~\cite{gargMobQoSMobilityAwareQoSDriven2019,gargSDNBasedSecurePrivacyPreserving2019}. 
The blockchain cannot support the more and more frequent transactions 
in the V2G network at a lower cost. 
Therefore, 
~\citet{hassijaBlockchainBasedFrameworkLightweight2020} proposed a lightweight 
blockchain-based protocol, 
called DAG-based V2G network (DV2G).

\subsection{Smart City} \label{subsec:dltcities}

The concept of cloud-based 
smart cities have already been evident~\cite{khanCloudBasedBig2015,khanFrameworkCloudbasedContextaware2014}. 
However, 
cloud technology cannot solve all the requirements of smart cities 
in terms of decentralization and transparency. 
Therefore, 
it is practical to introduce DLT to 
enable the technical feasibility of the anticipated smart cities.  

Smart cities aim to use information technology to integrate
various subsystems and services of the city, 
thereby optimize city orchestration system by
improving city governance and operating efficiency, 
and thus improving the quality of living. 
Smart cities could adopt DLT to enable public to participate in the 
city's decision-making process while ensuring a high degree of transparency. 
Based on our knowledge-graph analysis, 
we found that the following intrinsic characteristics attainable via DLT 
are most frequently mentioned in the smart city literature.

\begin{description}
	\item[Security] 
	Public data are to be trusted and shall never be maliciously tampered.
	
	\item[Transparency] Transactions should be highly transparent,
	and the necessary information should be accessible easily by the public.
	
	\item[Privacy] Public data need to be processed at several administrative 
	and geographic levels and should ensure strong privacy preservation.
\end{description}

Smart cities consist of a variety of IoT devices, sensors and actuators 
which are the main components of the infrastructure for improved intelligence. 
However, 
these devices are often fragile and vulnerable to data security breaches. 
Although~\citet{malikNonIntrusiveDeploymentBlockchain2019} reported that 
the inherent properties of DLT may be solved, 
these security challenges make the deployment process of 
a blockchain-based smart city extremely complicated.
Therefore, 
a systematic method is to be proposed to 
automate (and orchestrate) the deployment of DLTs over smart city environment. 

Research on citizen participation in urban decision-making processes has been ongoing, 
with the hope of adequately protecting citizen privacy at different levels.
~\citet{khanFrameworkCloudbasedContextaware2014} proposed 
a framework based on edge computing with blockchain. 
Blockchain in~\cite{khanFrameworkCloudbasedContextaware2014} is used to 
provide privacy protection, 
identity verification, 
and authorization functions in the transaction process in open urban governance. 
Edge computing is used for refined processing of different geographic 
and administrative levels of data to enhance privacy. 

\subsection{Smart Manufacturing and Industry 4.0} \label{subsubsec-dltmanufacture}

From the perspective of data, information, and knowledge flow, 
smart manufacturing perceives and obtains the real data of things through industrial IoT, 
and then extracts the knowledge contained in it to 
provide personalized industry 4.0 services~\cite{xi-fanWisdomManufacturingNew2014}.
These services include several other application-based requirements, 
including reducing delays in supplies and account collections. 
They may produce traceability throughout the supply chain, 
simplify the transaction links, 
and so on~\cite{mentzerDefiningSupplyChain2001,lambertIssuesSupplyChain2000}.
In this case, 
the inherent characteristics of DLT can meet such requirements of smart manufacturing, 
especially the supply chain logistics~\cite{wustYouNeedBlockchain2018}.
In our literature review of industry 4.0, 
the following characteristics are prevalent in our knowledge graph.

\begin{description}
	\item[Decentralization] All nodes jointly maintaining a ledger can always enhance the fault tolerance level of the entire network.
	\item[Trust] Each node in the manufacturing chain shall fully trust other 
	participants to record data honestly and maintain security.
	\item[Traceability] Manufacturing process 
	should be recorded timely and accurately as transactions, and could be used at any time for industry 4.0 automation and orchestration.
	\item[Transaction fee] Some middle boxes often appear in the manufacturing chain, 
	when there are no transaction fee (or low transaction fee) for between various entities of the system.
\end{description}

The advantages of blockchain traceability in the manufacturing industry 
have been widely studied in literature
~\cite{duranArchitectureEasyOnboarding2020}.
The composite material industry is a major supplier in many industries, 
especially semi-completed products used in the supply of regulated industries, 
ensuring quality and traceability.
~\citet{mondragonExploringApplicabilityBlockchain2018} has studied 
the ability of blockchain to be applied in the composite materials' industry, 
and pointed out which manufacturing process data should be recorded on the blockchain. 
The conclusion indicates that 
the blockchain can shorten the delivery time. 
The realization of product traceability will help reduce 
manufacturing costs and realize economies of scale in the future.

Researchers are also interested in how blockchain can help 
manufacturing companies improve their efficiency.
~\citet{koBlockchainTechnologyManufacturing2018} first reviewed the 
current application of blockchain in the financial industry and supply chain, 
and then explained the mechanism of blockchain to achieve real-time 
transparency and cost-saving in the manufacturing industry, 
and also established The theoretical model is to discuss the competition 
between manufacturing companies with and without blockchain in the case of a duopoly. 
The theoretical model shows that the blockchain helps to ensure the integrity of the distributed system. 
At the same time, 
through the consensus algorithm, 
it can Realize real-time transparency and save costs, 
thereby ensuring the sustainability of the manufacturing industry.

Another topic that has emerged with smart manufacturing is 
the \emph{machine-to-machine} (M2M) economy~\cite{raschendorferIOTAPotentialEnabler2019}. 
In the M2M economy, 
the machine can accept orders from customers, 
and then automatically order and pay for the required parts.
An order in M2M may require multiple production processes or parts, 
so the machine needs to divide the order into many sub-tasks, 
so the sub-tasks must be processed and paid quickly.
In this scenario, 
scalability and payment become the main constraints, 
so the blockchain may no longer be applicable.
~\citet{raschendorferIOTAPotentialEnabler2019} constructed a 
simple M2M economic system that accepts images uploaded 
by customers and orders raw materials (such as water and paint) 
through the machine for drawing.
All payments in the system are processed through the \emph{Tangle}, 
and IOTA's API is used to initiate payment requests automatically.
The results show that 
the DLT system based on DAG has great application prospects in the M2M economy, 
but it appears beneficial to 
consider the integration of smart contracts in the 
future.

\subsection{Intelligent Transportation and Autonomous vehicles} 
\label{subsubsec-dlttransport}

Intelligent transportation is an important part of smart cities, 
and a broader concept for the so-called \textit{smart mobility}.
Its purpose is to cope with the huge challenges brought by population growth, 
traffic growth and traffic congestion to cities, 
and to promote organized commuting process and transportation 
services through intelligence and automation~\cite{kumarSmartMobilityCrowdsourcing2018}.
Distributed scenarios such as car insurance, 
traffic control, and car sharing 
for intelligent transportation naturally require DLT to provide a better experience.
However, 
these scenarios contain a large amount of travel history data about the driver, 
so the application of DLT in intelligent transportation should pay attention to privacy.
In our literature review of intelligent transportation, 
the following characteristics are frequently mentioned.
\begin{description}
	\item[Privacy] Personal historical activity information is 
	protected and complies with relevant laws and regulations. 
	
	\item[Scalability] In the scenario of serving many users, 
	it can maintain good performance and low cost at the same time.
\end{description}

Vehicle insurance claims often require extra costs. 
In order to solve this problem,
~\citet{baderSmartContractBasedCar2018} designed a smart contract 
called \emph{CAIPY} based on \emph{Ethereum}, 
assuming that cars are equipped with a set of IoT devices that use DLT technology, 
and can reliably detect the conditions faced by the car 
(such as collisions or parts Failure) 
whether it meets the compensation requirements stipulated in the contract, 
so \emph{CAIPY} can automatically initiate a claim 
based on the feedback from the IoT device, 
and the insurance company can complete the subsequent transfer process 
only through the claim request.
At the same time, 
\emph{CAIPY} uses an insurance token to mitigate price fluctuations in Ethereum.

New research about the sharing economy in intelligent transportation 
has been evident.
Ride-sharing service is another good example of a distributed scenario, 
where the location of the vehicle and service can be traced and dispatched through DLT, 
the payment data can be transparent, 
and the end to end delivery of the passengers can be more trustworthy.
\citet{valastinBlockchainBasedCarSharing2019}, 
for instance, used Solidity programming 
language 
and Ethereum to develop a peer-to-peer car-sharing platform without central authorization, 
thereby preventing car owners from misusing customers' personal data and reducing costs. 
At the same time, 
the platform passed an order Brand service 
that incorporates both B2B and B2C scenarios into the scope of services.
~\citet{mladenovicCooperativeFrameworkUniversal2019} focused on 
the long-term cooperative behaviour of mobile credit and proposed a framework 
for decentralized control.
In order to complete the cooperation, 
the transaction system must have identity management functions and achieve long-term interaction consistently in a distributed fashion.

\section{Future Directions} \label{sec-discussion}

We discuss several potential DLT variants and impeding developments, 
to guide towards maximizing the utility of DLTs 
considering the dynamics of IoT networks.

\subsection{Virtualization} \label{subsubsec-virtualization}

As argued in~\Cref{sec-challenges}, 
it has been known that the flexible access of 
the DLT via the cloud platform are believed to be
the next step via its virtualization~\cite{yuVirtualizationDistributedLedger2018}. 
By virtualizing the underlying physical resources and 
encapsulating powerful DLT functionalities, 
scalability and flexibility of DLT can be attained. 
Many thanks to the customizable API, 
such virtualization often provides several DLT prototypes 
(e.g., blockchain, DAG etc.) at the same time.

On the one hand, 
virtualization provides major rethink on the direction of designing DLT, 
with a few recent important research works~\cite{yuServiceorientedBlockchainSystem2019,krishnaswamyMicroservicesBasedVirtualizedBlockchain2019}. 
The feasibility and performance of DLT virtualization requires further evaluations. 
On the other hand, 
while realizing on the DLT virtualization, 
we first need to solve critical problems, 
such as \textit{data ownership and centralization issues}. 
For example,
entities that are providing virtualized DLT services 
could become a new centralized node in DLT.

\subsection{Cross-Prototype Compatibility} \label{subsubsec-intergration}

As we have already discussed in~\Cref{subsec-prototype} and~\Cref{sec-dlts}, 
there are several prototypes of DLT. 
An efficient requirement in designing an optimal DLT over IoT 
is for transferring assets or values among different components.
At present, the assets of different DLT prototypes 
can be traded through a third-party platform, 
however, 
such a consequence has already raised a few concerns of centralization.
Therefore, 
scholars are committed to building a value transfer mechanism between different ledgers 
without relying on any intermediaries~\cite{thomasProtocolInterledgerPayments2015}.
Actually, 
there are already many solutions to transfer assets between different blockchains, 
which is called cross-chain payment, 
such as atomic cross-chain swaps~\cite{herlihyAtomicCrossChainSwaps2018}, 
side chains~\cite{backEnablingBlockchainInnovations2014}, bridging, 
payment channels~\cite{deckerFastScalablePayment2015}, ledger-of-ledgers approaches, etc.
~\citet{sirisInterledgerApproaches2019} have discussed and compared these methods in detail, 
For instance, 
ILP allows value to be transferred between all types of DLT prototypes, 
and is most likely to become an industry standard~\cite{hope-bailieInterledgerCreatingStandard2016}, 
but there are also concerns about centralization because of the dependence on third parties~\cite{koensAssessingInteroperabilitySolutions2019}. 
Most of these are to be used for blockchain, 
other methods have a smaller scope of application 
and have their own unique advantages and limitations depending on the scenarios. 

There is no doubt that different DLT prototypes 
will greatly promote the value flow and stimulate the prosperity of DLT. 
However, 
most of the current solutions focus on the value transfer between blockchains, 
It seems that the asset flows between different DLT prototypes 
still need to rely on the third party to build trust. 
Therefore, 
establishing a cross-prototype value transfer mechanism and making them compatible with each other with 
low complexity and cost are still impeding research problems in this field.

\subsection{IoT and Decentralized Edge Computing} \label{subsubsec-consensusinn}

The proposed integration of DLT with IoT faces resource constraints 
such as computing power, 
as well as performance concerns such as network latency. 
But the rise of edge computing and deployments of 5G could alleviate these problems.
The combination of IoT and edge computing has been fully demonstrated
~\cite{fernandoMobileCloudComputing2013}. 
At the same time, 
edge computing and DLT could benefit from each other
~\cite{queraltaBlockchainMobileEdge2021,isajaCombiningEdgeComputing2017}.
Edge computing always relies 
on data centers~\cite{vargheseChallengesOpportunitiesEdge2016}, 
so its infrastructure poses high communication bandwidth and sufficient computing resources, 
such as GPUs or super-computing processors~\cite{liuSurveyEdgeComputing2019}, 
which are beneficial to alleviate the resource limitations for DLT. 
In addition, 
the edge computing servers in the same data center 
have extremely low latency (optical fibre connections), 
which may create a new prioritized data flow for DLT traffic and 
reduces the regular data flow in the whole core network 
by a small factor~\cite{yuSurveyEdgeComputing2018}.

Merging DLT, IoT, and edge computing would inspire incredible innovations. 
However,  
the edge computing infrastructure is usually 
highly distributed~\cite{shiEdgeComputingVision2016}, 
which means that 
it is difficult for developers to 
guarantee a consistent experience for different regions. 

\subsection{Empowering with Federated Learning} \label{subsubsec-fl}

DLT, IoT, and FL have a few common characteristics and 
as mentioned all of them are distributed by nature
~\cite{jainCloudEdgeDistributed2017,bonawitzFederatedLearningScale2019}. 
They may naturally fit together, 
and the proposed convergence being the most probable 
high impact design for future systems. 
In fact, 
there have been a lot of studies considering the use of FL and DLT over IoT
~\cite{pokhrelFederatedLearningBlockchain2020, 
	nguyenDIoTFederatedSelflearning2019, wuPersonalizedFederatedLearning2020}. 
However, 
FL faces the risk of privacy attacks~\cite{mothukuriSurveySecurityPrivacy2021}. 
Generally speaking, 
malicious attackers upload malicious gradient information 
to achieve member inference attacks~\cite{nasrComprehensivePrivacyAnalysis2019}. 
The invariance of DLT can bring a new approach to 
privacy protection in FL~\cite{quDecentralizedPrivacyUsing2020}.
For example,
~\citet{luBlockchainFederatedLearning2020} have built a blockchained FL method 
in the industrial IoT scenario, to better protect privacy, 
and~\citet{zhaoPrivacyPreservingBlockchainBasedFederated2021} use blockchain to replace 
the centralized aggregator in traditional federated learning to help home appliance manufacturers 
get feedback from users while protecting privacy.

In particular, 
the intrinsic distributed characteristics of the three concepts bring 
opportunities for their convergence~\cite{quBlockchainedFederatedLearning2021}. 
The related research is still in its infancy, 
so there are several opportunities. 
However, 
a very challenging problem is how to solve the overhead 
and delay that blockchain may bring. 

\subsection{Energy Efficient Consensus} \label{subsubsec-consensus innovation}

Considering its wide applicability for energy trading, 
DLTs have the potential to 
play a key role in impending natural concerns 
such as carbon emissions and energy trading platforms (smart grids)
~\cite{brilliantovaBlockchainFutureEnergy2019,khaqqiIncorporatingSellerBuyer2018,panApplicationBlockchainCarbon2019}. 
These studies fully investigate the potential contribution of blockchain to 
emissions trading schemes (ETS), 
such as helping parties build trust without intermediaries.
But one of the aspects that have often been overlooked is that DLT itself 
has its own energy and carbon emissions problems caused by the computationally intensive consensus mechanisms. 
A few researchers~\cite{trubyDecarbonizingBitcoinLaw2018,imbaultGreenBlockchainManaging2017,sedlmeirEnergyConsumptionBlockchain2020} 
have already noticed the possible impact of DLT 
on the environment and proposed enhanced solutions from the regulatory perspective~\cite{jiangPolicyAssessmentsCarbon2021}, 
however, 
an in-depth understanding of such an issue lacks in literature~\cite{nairApproachMinimizeEnergy2020}.

One may consider developing a lightweight consensus mechanism 
for minimizing the computing power requirements, 
however, such a design requires further investigations on the 
impact of the environment and their trade-off with underlying IoT network settings. 
The energy-saving in blockchain-enabled IoT networks has been 
an inevitable requirement for the development of DLT.  
The wide-scale deployments of IoT have been demanding for a 
more scalable and resource-optimal DLT consensus mechanisms. 
Therefore, 
our redesigns should focus on 
improving the underlying mechanism of PoW, PoS, and PoA, 
or rethink new consensus mechanisms suitable for IoT. 
This is reflected by the ongoing advances such as 
developments of \emph{Tangle}~\cite{popovTangle2018}, 
GaG~\cite{bairdSwirldsHashgraphConsensus2016} 
and Adaptive PBFT~\cite{misicAdaptingPBFTUse2021}.

In the future,  
the consensus mechanisms of DLTs  are to 
be co-designed with the underlying IoT requirements by
considering high scalability and network resource optimal approach.  
However,  
due to the inevitable nature of scalability and security trade-offs in IoT, 
it is delicate to balance the scalability and security, 
and the trade-off requires further investigations.

\section{Towards A New DLT-IoT Convergence Architecture} \label{sec-architecture}

With relevant insights from the earlier sections and analyss, 
in this section, we introduce a high level framework of DLT-IoT convergence by considering a holistic solution background for resolving
the impeding challenges of DLTs over IoT applications.
Existing DLT architecture consists of a four-layer approach~\cite{xuInternetThingsIndustries2014,atlamBlockchainInternetThings2018}, 
viz. \emph{data layer}, \emph{network layer}, 
\emph{consensus layer}, \emph{function layer} and \emph{application layer}. 
To this end, 
we aim to enhance the existing features of the developed architectures, 
to integrate DLT for IoT applications tightly. 
As mentioned, 
IoT comprises thousands of sensing devices, 
which will serve as distributed nodes of the DLT network, 
and ``DLT over IoT'' is therefore subject to resource constraints. 
Unfortunately, 
Almost no existing architecture shows such intrinsic capabilities and
hierarchical relationship with limitations for the DLT over IoT devices. 
In this survey, in sharp contrast to existing architectures,
we introduce a \emph{hardware layer} at the bottom to quantify the impact 
of implementing DLT over IoT due to limited resources and computing constraints. 

On the data layer, 
these architectures often only distinguish between blockchain and DAG. 
However, 
more prototypes have emerged recently. 
These prototypes differ in the data structure, 
represent different node organization methods, 
and even represent an integration method. 
Therefore, 
it is not appropriate to use the \emph{data layer} to describe different DLTs. 
We use the \emph{prototype layer} alone to introduce almost all DLT prototypes. 
Correspondingly, 
more consensus mechanisms will be included in the \emph{consensus layer}. 
Finally, 
we designed a new \emph{function Layer} to show the close coupling between 
DLT characteristics and industry applications.

\begin{figure}[t]
	\centering
	\includegraphics[scale=0.4]{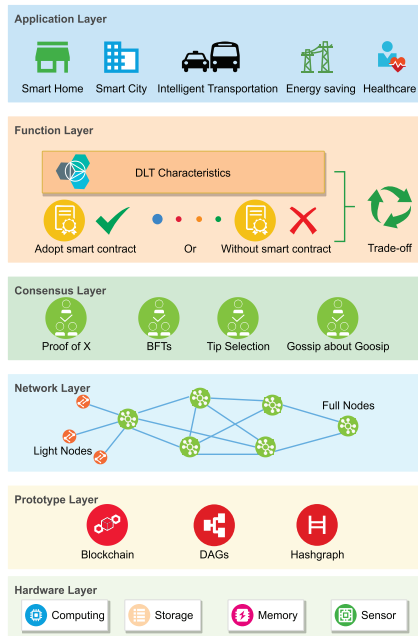}
	\caption{Proposed Architecture of DLT for IoT}
	\label{fig:architecture}
	\vspace{-5 mm} \end{figure}

\Cref{fig:architecture} shows the organization of this paper. 
Starting from IoT devices (hardware layer), 
we discussed the DLT prototypes that can be implemented (prototype layer), 
how nodes are connected into a network (network layer), and how nodes in the network reach an agreement (consensus layer), 
which parts need to be trade-off before application (function Layer), 
and different industry application scenarios (application layer). 

\subsubsection{Hardware Layer} \label{subsec-Hardware}

IoT consists of many sensing devices. 
The potential of the combination of DLT and IoT is 
derived from similar distributed scenarios. 
These hardware devices will become nodes of the DLT network 
and carry some functions of DLT. 
In short, 
Under normal circumstances, 
hardware devices will be the foundation of IoT and DLT. 

\begin{description}
	\item[Power] 
	There are many types of power supplies for IoT devices, 
	such as dry batteries, button batteries, solar and AC power. 
	Regardless of the type of power supply, 
	it is impossible to provide a near-infinite energy supply to IoT devices. 
	This is also unrealistic due to volume and cost considerations. 
	Therefore, 
	the power of IoT devices is as small as possible.
	
	\item[Computation] 
	IoT devices are usually developed based on single-chip microcomputers. 
	The CPU used is a low-power processor suitable for mobile terminals. 
	Its computing power is far lower than common desktop processors, 
	and even some smartphones, 
	which only handle some simple and necessary computing tasks.
	
	\item[Storage] 
	IoT storage is usually used to store some buffered data, 
	so the storage space is minimal. 
	The storage space of RAM is often at the level of tens to hundreds of KB, 
	while the storage space of Flash is often at the level of several MB. 
	Some IoT devices may be equipped with external storage due to market needs, 
	such as smart cameras, 
	but this is very rare for IoT devices and will bring very high costs.
	
	\item[Wireless] 
	There are many types of wireless communication protocols for IoT devices, 
	but most of them use short-range communication protocols. 
	Common ones are NFC, Zigbee, and Bluetooth. 
	Although there are differences between them, 
	the bandwidth is not high, 
	so it is difficult to transmit large files. 
	WiFi can achieve high bandwidth, 
	but the cost and delay are relatively high. 
	5g has brought a revolutionary change to IoT. 
	5g can reduce the delay by several milliseconds, 
	and has a very high bandwidth, 
	but it is still in the construction stage, 
	and the cost is high, 
	so it is suitable for application in the industrial Internet of things. 
	
	\item[Sensor] 
	Sensing equipment gives the IoT the ability to perceive the world. 
	Different scenarios require different sensors, 
	such as infrared sensors, GPS, and temperature sensors.
\end{description}

\subsubsection{Prototype Layer} \label{subsec-prototype}

Recently,
the blockchain came into focus with the popularity of BTC
~\cite{nakamotoBitcoinPeertopeerElectronic2008}.
Due to the great success in finance and cryptocurrency
~\cite{tschorschBitcoinTechnicalSurvey2016}, 
blockchain has attracted significant interest from researchers and practitioners.
After years of development, 
blockchain has become the most representative DLT prototype and the first 
choice when using DLT in IoT scenarios~\cite{sangIEEESABlockchain2020}. 

\begin{figure}[htbp]
	\centering
	\includegraphics[width=0.7\textwidth]{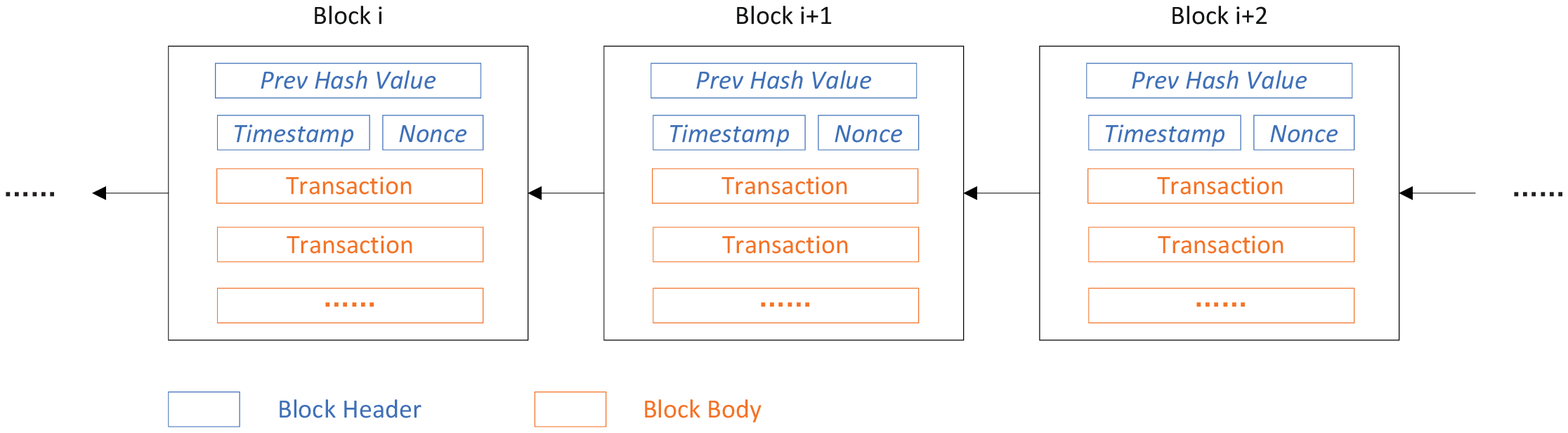}
	\caption{The structure of the blockchain}
	\label{fig:blockchain}
	\vspace{-5 mm} 
\end{figure}

The structure of the blockchain is shown in the \Cref{fig:blockchain}. 
Multiple transactions in the blockchain are verified 
by the verification node and packaged into a transaction set, 
that is, 
a block, 
and the first block is called the genesis block. 
To ensure the \emph{immutability} of the data, 
the block also contains a timestamp and the hash value of the previous block. 
In cryptography, 
the hash value is unique, 
and any change of the content will change the corresponding hash value. 
If other nodes in the network can verify the hash value by the nonce, 
that is, 
verify the validity of the block, 
then other nodes will accept the block. 
This process is called the consensus mechanism. 
To be exact, 
the consensus mechanism is a set of procedures and rules to 
ensure the consistency of the accounts of each participating node~\cite{swansonConsensusasaserviceBriefReport2015}.

Blockchain adds blocks in a chronological order 
and uses PoW as a consensus mechanism.
However, 
they manifest scalability and resource consumption issues. 
Therefore, 
some new DLT mechanisms have been proposed, 
such as \emph{IOTA} and \emph{Hashgraph}. 
They are designed for IoT, 
use DAG as the network topology, 
and claim to overcome these shortcomings of the blockchain. 
We will discuss and compare these prototypes in detail in \Cref{sec-dlts}.

\subsubsection{Network Layer} \label{subsec-network}

Due to decentralization, 
there is no right priority node in the DLT network, 
but each node is in an equal position, 
so it is a kind of peer-to-peer network (P2P). 
Even if there is no master-slave relationship 
between nodes in the network, 
due to some considerations, 
such as cost, 
their roles in the network, 
in other words, 
their functions may be different.
For example, 
in the tangle network of \emph{IOTA},
\emph{Full Node} stores all transactions since the last snapshot of the network, 
\emph{Lightnodes} do not store the transactions, 
but access the network functionality through full nodes, 
and \emph{Permanodes} are data storage nodes 
that do not delete data in a snapshot and 
make this data available to other nodes in the network. 
It is worth noting that 
the network layer does not only mean the network topology composing of nodes, 
but also the communication links between nodes.

\subsubsection{Consensus Layer} \label{subsec-consensus}

As mentioned earlier, 
all nodes in the DLT network maintain the same ledger under 
the premise of honest sending and recording data. 
However, 
in reality, 
many factors make this premise unable to be guaranteed, 
such as network delay, network failure, and malicious nodes, 
which will lead to differences between nodes on the effectiveness of the ledger. 
Therefore, 
we must rely on a complete set of rules and methods to 
avoid malicious nodes' information being added to the general ledger, 
and ultimately ensure that all nodes in the network 
agreement on the correctness of the ledger.

The current consensus mechanism mainly includes \emph{Proof of `X'}, 
Byzantine fault tolerance, Tip Selection algorithm, and Gossip about Gossip. 
The first two are mainly used in blockchain, 
while the latter two are consensus mechanisms 
adopted by \emph{Tangle} and \emph{Hashgraph}, respectively.

\begin{description}
	\item[Proof of `X'] 
	If the probability that 
	a node obtains the right to keep accounts is related to 
	the proportion of resources that are difficult to monopolize by the node, 
	and there is an algorithm that can quickly verify that a node does own this part of the resources, 
	this type of consensus mechanism is called \emph{Proof of `X'} consensus. 
	X Represents the resources required to compete for the right to bookkeeping. 
	This kind of consensus mechanism includes PoW that competes with computing power, 
	PoS that compete with rights and interests, 
	and PoA, PoSV, Casper, etc. that improve some defects of PoS, 
	and PoST and Proof of Authority that allocates accounting rights 
	based on other resources that are not easily monopolized.
	
	In the PoW consensus mechanism, 
	this resource is the so-called ``computing power", 
	that is, 
	a certain amount of CPU or GPU computing time. 
	The hash process of PoW can be expressed by \cref{eq1},
	\begin{equation}\label{eq1}
		H(nonce + Block) < Target
	\end{equation}
	that is, 
	the verification node constantly changes the nonce of the blockhead 
	to make the hash value of the block meet the \emph{Target} condition.
	\emph{Target} is the same for all nodes, 
	so the probability of obtaining accounting rights depends on the speed of computing hash, 
	and to ensure that the time expectation of mining blocks in the network is fixed, 
	the difficulty of the target will be adjusted according to a specified algorithm.
	Most blockchain prototypes such as BTC and litecoin use the PoW consensus mechanism, 
	but there are differences in the hashing algorithms and incentive mechanisms used.
	
	The computing power competition of PoW brings a lot of energy consumption, 
	so PoS allocates accounting rights according to the proportion of 
	certain rights and interests held by nodes in all nodes of the network, 
	to achieve no less security than the PoW consensus mechanism and 
	avoid energy consumption at the same time. 
	Peercoin puts forward the concept of ``coin age.'' 
	Token coin age is equal to the number of tokens multiplied by the time 
	elapsed since the last transaction of this part of the token. 
	The \emph{Unspent TX Output} contains a certain number of tokens, 
	and records the mining time of the block containing the transaction, 
	so that the corresponding coin age can be calculated. 
	When UTXO was used, 
	the coin age of this part of the token was also cleared. 
	Peercoin uses coinday as the unit of coin age 
	(1-day accumulated coin age of 1-unit token). 
	The node can take its own UTXO as the core input 
	and obtain the right of production block by consuming coin age. 
	The consensus process of PoS can be expressed by formula \cref{eq2}.
	\begin{equation}\label{eq2}
		H ( \text { StakeModifier } + \text { Timestamp } ) < \text { 
			BaseTarget } * \text { CoinAge}
	\end{equation}
	
	Although PoS reduces energy consumption, 
	its ability to fight hard forks is weak, 
	so improved PoS consensus mechanisms have emerged, 
	such as Proof of Activity, Proof of Stake Velocity, and Casper of Ethereum. 
	In addition, 
	it should be pointed out that \emph{Proof of `X'} reached an indirect consensus, 
	because the content of a block is determined by the node that produces it, 
	and then verified by other nodes.
	
	\item[BFTs] 
	Different with \emph{Proof of `X'}, 
	BFTs consensus mechanism adopts the principle that the minority is subordinate 
	to the majority to vote directly on the content of the block, 
	The representative ones are Practical Byzantine Fault Tolerance (PBFT) 
	and Federal Byzantine Fault Tolerance (FBT). 
	
	\emph{Hyperledger Fabric} use PBFT as its consensus mechanism 
	and introduces the concept of ``View'' and ``Replica''. 
	The replica includes primary nodes and backup nodes. 
	The primary node is usually selected randomly or alternately at 
	the beginning of each round of the consensus process. 
	View represents the process of a master node distributing a request. 
	When a round of consensus begins, 
	the effectiveness of the master node is first checked. 
	If the backup node detects the failure of the primary node, 
	it needs to elect a new primary node, 
	which is called ``viewchange". 
	The consensus process is divided into pre-preparation, 
	preparation, and confirmation stages, 
	also known as the ``three-phase agreement". 
	PBFT can provide fault tolerance of $ f $ for a 
	blockchain network composed of $ n = 3f + 1 $ nodes, 
	and the fault tolerance is approximately $ 1/3 $. 
	Since the number of communication between nodes in a round of consensus is proportional to $ n^2 $, 
	considering the existence of communication delay, 
	when the number of nodes reaches a certain scale, 
	the efficiency of PBFT will be severely reduced.
	
	FBA is another consensus mechanism to 
	solve the problem of \emph{Byzantine Generals}. 
	It forms a trust, Federation, in the subnetwork 
	and regards the subnetwork as a whole as the network node. 
	Therefore,
	the minimum connectivity between the sub-networks should also be guaranteed. 
	Specifically, 
	the nodes in FBA can be divided into verification nodes participating 
	in the consensus process and the non-verification nodes not participating. 
	The service nodes only accept the votes from the verification nodes and 
	trust that they will not cheat jointly. 
	In this way, 
	the service nodes and their trusted verification nodes form a federation.
	
	\item[Tangle] 
	\emph{Tangle} is the consensus mechanism of \emph{IOTA}. 
	All transactions in the \emph{Tangle} are called ``Tips''. 
	When a node initiates a new transaction, 
	it needs to use the \emph{Tip Selection algorithm} to 
	select the previous two tips to verify, 
	and point the new transaction to these two transactions, 
	thus forming a structure of the directed acyclic graph, 
	which allows the verification of tangle transactions to be carried out in parallel, 
	and with the continuous development of the tangle network, 
	the speed of confirmation becomes faster and faster.
	It is worth noting that to add a new transaction to \emph{Tangle}, 
	nodes must also perform PoW calculation, 
	but the difficulty is much lower than that in the blockchain, 
	and the purpose is to prevent malicious attacks.
	
	\item[Gossip about Gossip] 
	\emph{Hashgraph} uses the gossip protocol, 
	which refers to information owned by one node but not owned by other nodes. 
	In \emph{Hashgraph}, 
	each node will broadcast its own information to the neighbouring nodes randomly. 
	The neighbouring nodes will package the new transaction information 
	and their own transaction information into an event, 
	which is similar to the block, 
	and broadcast it again, 
	so it is called the gossip about gossip (GaG).
	In this way, 
	if a single node owns the new transaction information, 
	it will spread to the whole network exponentially until each node owns this transaction. 
	At the same time, 
	\emph{Hashgraph} uses a method called \emph{virtual voting} 
	~\cite{bairdSwirldsHashgraphConsensus2016}, 
	which makes each node execute the voting algorithm independently, 
	and will eventually get the same result.
	Compared with PBFT, the gossip of gossip reduces the communication complexity to 
	$ O(n) $.
	
\end{description}

\Cref{tab:consensus} shows the comparison of different consensus mechanisms, 
where the security boundary refers to the computing power required to 
successfully implement malicious attacks.

\begin{table}[htbp]
	\centering
	\caption{Comparison of different consensus mechanisms}
	\begin{tabular}{lccccc}
		\toprule
		& \textbf{PoW} & \textbf{PoS} & \textbf{Tangle} & \textbf{BFTs} & 
		\textbf{GaG} \\
		\midrule
		Prototype & BTC   & Ethereum & IOTA & Hyperledger & Hashgraph \\
		Latency~\cite{zamaniRapidChainScalingBlockchain2018, karameSecurityScalabilityBitcoin2016, bairdHederaGoverningCouncil2018} & High  & Low & Most Low & Low & Low \\
		Throughput~\cite{popovTangle2018,kimMeasuringEthereumNetwork2018} & Most Low & High  & Most High & High & High \\
		Energy Saving~\cite{devriesBitcoinGrowingEnergy2018} & No & Yes   & Yes   & Yes   & Yes \\
		Scalability~\cite{poonBitcoinLightningNetwork2016, pongnumkulPerformanceAnalysisPrivate2017, popovTangle2018} & Good & Good & Best & Poor & Best \\
		Transaction Fee~\cite{saadMempoolOptimizationDefending2019} & High & Low   & No Fee & Low & Low \\
		\cmidrule{2-4}
		Fault Tolerance~\cite{sajanaBlockchainApplicationsHyperledger2018,popovTangle2018,bairdHederaGoverningCouncil2018} & \multicolumn{3}{c}{$ \SI{50}{\percent} $} & $ \SI{33}{\percent} $~ & $ \SI{33}{\percent} $ \\
		\bottomrule
	\end{tabular}%
	\label{tab:consensus}%
\end{table}%

\subsection{Function Layer} \label{subsec-tradeoff}

DLT has so many characteristics, 
and has a complex relationship with applications. 
The fact is that it is impossible to adopt all DLT 
characteristics in a certain application scenario. 
Instead, 
it is necessary to make trade-offs to ensure that 
the key requirements are met and choose DLT prototypes accordingly, 
and before that, 
a comprehensive summary of the characteristics of DLT must be made. 
Different from the existing work, 
we try to use more objective methods to replace the 
literature survey and expert interviews, 
so the knowledge graph has become the first choice. 

\begin{figure}[htbp]
	\centering
	\includegraphics[width=0.68\textwidth]{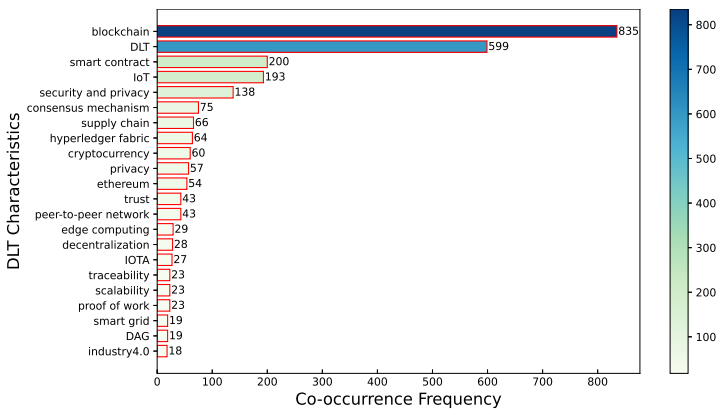}
	\caption{Knowledge Graph: Extract node labels based on the frequency}
	\label{fig:characteristics}
	\vspace{-5 mm} \end{figure}

As shown in the \Cref{fig:characteristics}, 
we used \emph{Citespace} to generate a keyword co-occurrence network. 
Since we are only interested in the DLT characteristics, 
we no longer use the timeline view. 
At the same time, 
we use frequency as the basis for selecting node labels. 
Therefore, 
the larger the node, 
the higher the co-occurrence frequency of the corresponding label in the citation.
To make it easier to observe, 
we have hidden other keywords that are not related to the characteristics, 
thereby determined the following DLT characteristics.

\begin{description}
	\item[Distributed] 
	The DLT prototype consists of many computer nodes. 
	Normally, 
	the nodes are geographically distributed in various locations, 
	and they work together to complete common tasks~\cite{baranDistributedCommunicationsNetworks1964}.
	
	\item[Decentralization] 
	There is no central node in the node network, 
	but all or part of the nodes reach 
	an agreement through consensus mechanism~\cite{bencicDistributedLedgerTechnology2018}.
	
	\item[{Peer-to-Peer Network}] 
	The nodes in the DLT prototype share resources through the network and 
	can be directly accessed by other nodes without passing through intermediate entities.
	
	\item[Immutable] 
	The nodes faithfully record the data and reach an agreement 
	on the validity of the data 
	through the consensus mechanism. 
	Fake data and malicious tampering will eventually be rejected.
	~\cite{magazzeniValidationVerificationSmart2017}.
	
	\item[Cryptography] 
	DLT uses cryptography to verify whether the data has been modified 
	and the identity of both parties to the transaction, 
	thereby ensuring the security of the transaction and data. 
	The most widely used technologies are hash functions and asymmetric encryption algorithms.
	~\cite{anjumBlockchainStandardsCompliance2017,dunphyFirstLookIdentity2018}.
	
	\item[Transparency] 
	DLT requires nodes to join or exit freely, 
	and all data can be accessed by as many nodes 
	as possible~\cite{sankarSurveyConsensusProtocols2017}.
	
	\item[Anonymity] 
	The nodes in DLT exchange data through pre-made rules 
	and their respective network addresses 
	without revealing their true personal identities~\cite{kuzunoBlockchainExplorerAnalytical2017}.
	
	\item[Scalability] 
	Due to the large number of nodes participating, 
	DLT requires new nodes to join continuously without causing congestion in 
	the entire system~\cite{zhangDependableScalablePervasive2018}. 
	If the scalability of a DLT prototype is insufficient, 
	it will face the upper limit of performance.
	
	\item[Throughput] 
	The ability of the DLT system to process transactions per unit time is called throughput. 
	The higher the throughput, 
	the higher the application value of the system.
	
	\item[Latency] 
	The time from initiation to confirmation of a new transaction 
	is called the latency of the DLT system. 
	If the latency is high, 
	to obtain the priority of transaction confirmation, 
	the node has to pay some fees for the verification node, 
	which will increase the cost.	
\end{description}

As the conceptual boundaries of some characteristics overlap, 
we further summarize the above characteristics into three categories: 
decentralization, privacy, and performance, 
to better describe the trade-off relationship among DLT characteristics.
We summarize them in \Cref*{tab:dltcharacter} (
discussed earlier in \Cref{subsec-challenginapplication}).

\begin{table}[t]
	\centering
	\caption{Summary of DLT Characteristics}
	\begin{tabular}{lll}
		\toprule
		\textbf{Decentralization} & \textbf{Privacy} & \textbf{Performance} \\
		\midrule
		Distributed & Immutable & Scalability \\
		Peer-to-Peer Network & Cryptography & Throughput \\
		Transparency & Anonymity & Latency \\
		\bottomrule
	\end{tabular}%
	\label{tab:dltcharacter}%
\end{table}%

\subsection{Application Layer} \label{subsec-applayer}

As mentioned earlier, 
DLT-IoT convergence is a bigger arrangement, 
and there are already many scenarios showing the exciting results of this convergence, 
such as smart home, smart health, smart energy, smart manufacturing, and so on. 
IoT, together with other advanced technologies, 
will comprehensively transform life and manufacturing, 
and eventually, make the society move towards the industrial 4.0 era, 
and DLT will become a key link in this process. 
However, 
in practical applications, 
a thorny problem is how to make an appropriate trade-off between the 
characteristics of DLT, 
to give full play to the greatest advantages of DLT, 
so we have discussed the use case of DLT-IoT convergence 
in detail in \Cref{sec-usecases}, 
and try to provide clear guidance for the handling of this coupling.

\section{Conclusion} \label{sec-conclusions}

We have developed a comprehensive survey to present a holistic view of 
the benefits and challenges of the convergence of DLT with IoT. 
Our systematic review process employed an extended knowledge graph approach 
to develop a pragmatic understanding of the evolution of DLT 
while considering IoT requirements. 
Furthermore, we elaborated several dimensions on the real DLT-IoT use cases scenarios. 
We also focus on the specialized convergence of DLT with IoT for
the automation and orchestration of the applications. 
We found that 
such a convergent IoT-native design of DLT enables several facets of IoT industry verticals 
and significantly outperforms the existing loosely coupled DLT designs.

\bibliography{yourbib.bib}
\bibliographystyle{ACM-Reference-Format}


\end{document}